\documentclass[%
 aip,
 sd,%
 amsmath,amssymb,
preprint,%
]{revtex4-1}

\usepackage{lineno,hyperref}
\usepackage{bm}
\usepackage{graphicx}
\usepackage{subcaption}
\modulolinenumbers[5]

\begin{document}
%
%
%

\title{Investigation of surface boundary conditions for continuum modeling of RF plasmas}

\author{A. Wilson}
\email{agw0001@uah.edu.}
\author{B. Shotorban}%
 \email{babak.shotorban@uah.edu.}
\affiliation{ 
Department of Mechanical and Aerospace Engineering, 
The University of Alabama in Huntsville, Huntsville, AL 35899
}%

\date{\today}
\begin{abstract}
This work was motivated by a lacking general consensus in the exact form of the boundary conditions (BCs) required on the solid surfaces for the continuum modeling of RF plasmas. Various  kinds of number and energy density BCs on solid surfaces were surveyed  and  how they interacted with the electric potential BC to affect the plasma was examined in two fundamental RF plasma reactor configurations.  A second-order local mean energy approximation with equations governing the electron and ion number densities, and the electron energy density was used to model the plasmas. Zero densities and various combinations of drift, diffusion and thermal fluxes were considered to set up BCs. It was shown  that the choice of BC can have a significant impact on the sheath and bulk plasma. The thermal and diffusion fluxes to the surface were found to be important. A pure drift BC for dielectric walls failed to produce a sheath.
%
\end{abstract}

\maketitle


\section{\label{sec:Intro}Introduction}  

Radiofrequency (RF) plasmas are encountered in many situations of practical interest such as manufacturing processes and laboratory experiments.  RF plasmas are used for etching and deposition of thin films on semiconductors, plasma enhanced chemical vapor deposition (PECVD) \cite{Stoffel1996}, producing quantum dots \cite{Michalet2005}, plasma synthesis \cite{Pi2008}, coating nanoparticles \cite{Cao2002}, and producing carbon nanotubes \cite{Meyyappan2003}.  They are also used  in some dusty plasma experiments \cite{Shukla2001} where phenomena such as Coulomb crystallization \cite{Ikezi1986,Schweigert1996} and dust charging \cite{Matthews2011,Havnes1990} are examined and in plasma medicine applications ranging  from equipment sterilization \cite{Moisan2001,Laroussi2005} to wound healing \cite{Fridman2008} and possibly cancer treatment \cite{Schlegel2013}. Modeling  through continuum (hydrodynamic) approaches has been an essential tool in understanding RF plasmas in both basic and applied research setups \cite{Park1990,Goedheer2004,Parent2014,Zakari2015}. 

In the continuum modeling of plasmas, partial differential equations derived from the first two or three moments of the Boltzmann equation are solved. There are two common continuum models for RF plasmas. One is  based on the local field approximation where a set of drift-diffusion equations describing the time and space variations of the ion and electron number densities are solved \cite{Barnes1987,Grubert2009}. The other is based on the local mean energy approximation where  an additional drift-diffusion equation is solved for the electron energy  \cite{Graves1986,Gogolides1992,Hammond2002}. In both approaches, a poisson equation is solved for the electric potential. 

Although boundary conditions (BC's) are essential for continuum modeling,  what constitutes an adequate BC on the solid surfaces of an RF plasma  is not completely described in the literature. Continuum models for RF plasmas conventionally use the same boundary conditions as DC plasmas.  There has been some detailed studies focused on the BC's  for direct current glow discharges \cite{Hagelaar2000BC,Surzhikov2004}; however, the competence of these conditions for RF plasmas has not been thoroughly investigated. 
In general, RF plasma continuum models use the local-mean-energy approximation which requires solving the electron energy equation. In contrast, DC plasmas often use the local-field-approximation which neglects the electron energy equation, hence the energy boundary conditions are not included  \cite{Hagelaar2000BC,Surzhikov2004}. Moreover, DC plasma boundary conditions often include the secondary electron emission (SEE) due to ion impact on the surfaces \cite{Hagelaar2000BC,Bouanaka2013} whereas RF plasmas often neglect it.

A continuum plasma model requires a set of BC's for the number and energy densities, and a BC for the electric potential. For the electric potential, surfaces are often grounded or have a known voltage, but could instead be dielectrics \cite{Boeuf1995,Davoudabadi2007} where the surface develops a non-uniform charge as a result of the current from the plasma. The number and energy densities are set to zero at the surface in some  studies \cite{Passchier1993,Davoudabadi2009} on the assumption that the charged particles are absorbed by the surface. 
In  other studies, the flux to the surface is specified based on the drift toward the boundary \cite{Davoudabadi2007,Horn2011} and the thermal motion \cite{Boeuf1995,Paunska2011}. 
The SEE effects at boundaries are often neglected in RF discharges although occasionally they are included \cite{Gozadinos2003}. 

In the  works reviewed above, the rationale for selecting the BC kind is rarely provided and the importance of the choice is  not fully discussed or quantified. The current work examines and compares various boundary conditions  for the number and energy densities and determines their effects in two fundamental RF plasma setups. The differential equations of the continuum model used here are presented in Sec. \ref{sec:PlasmaEqs}. 
The details of the boundary conditions are discussed in Sec. \ref{sec:BCeqs}. The numerical methods used to solve the system of equations are illustrated in Sec. \ref{sec:NumMethod}. 
The results are discussed in Sec. \ref{sec:Results} and conclusions are made in Sec. \ref{sec:Conclusions}.

\section{\label{sec:PlasmaEqs}Governing Equations}  

The second-order `local mean energy' model  \cite{Gogolides1992} is used for the RF plasmas studied in this work.  In this model, the equations governing the electron number density $n_e$, ion number density $n_i$, and electron energy density $\omega _e$ are:
\begin{equation} \label{eq:ne,i}
\frac{\partial n_{e(i)}}{\partial t} +\bm{\nabla}\cdot\bm{\Gamma}_{e(i)}=S_{e(i)} ,
\end{equation}
\begin{equation} \label{eq:EnergyTransport}
\frac{\partial\omega _e}{\partial t}+\bm{\nabla}\cdot\bm{\Gamma}_\omega =-e\bm{\Gamma}_e\cdot\bm{E}+S_\omega ,
\end{equation}
where
\begin{equation} \label{eq:neFlux}
\bm{\Gamma}_{e(i)}=\text{sgn}(q)n_{e(i)}\mu_{e(i)}\bm{E}-D_{e(i)}\bm{\nabla}n_{e(i)}  ,
\end{equation}
\begin{equation} \label{eq:EnergyFlux}
\bm{\Gamma}_\omega =\frac{5}{3}\left(-\omega_e\mu_e\bm{E}-D_e\bm{\nabla}\omega_e\right) ,
\end{equation}
and $\text{sgn}(q)$ is 1 for ions and -1 for electrons and electron energy.
Here,  $\bm{E}$ is the electric field calculated by:
\begin{equation} \label{eq:E_inst}
\bm{E}=-\bm{\nabla}\phi ,
\end{equation}
where $\phi$ is the electric potential which satisfies Poisson's equation:
\begin{equation} \label{eq:Poissons}
\nabla^2\phi=\frac{e}{\epsilon_0}\left( n_e-n_i\right),
\end{equation}

\noindent
where $e$ is the electron charge. In the equations above, $\mu_{e(i)}$ is the electron (ion) mobility, $D_{e(i)}$ is the electron (ion) diffusion coefficient and $\bm{E}$ is the electric field. Eqs. (\ref{eq:neFlux}-\ref{eq:EnergyFlux}) define the fluxes of  electrons (ions) and energy, respectively. 
The source term $S_{e(i)}$ in eq. (\ref{eq:ne,i}) accounts for the electrons and ions created by ionization. The gas is assumed to be singly ionized, therefore $S_i=S_e=k_in_en_\text{gas}$ where $k_i$ is the ionization rate coefficient. The ionization rate was determined by BOLSIG+ \cite{Hagelaar2005} which solves the electron Boltzmann equation and tabulates the ionization rate and excitation rates as a function of the mean electron energy. 
The mean electron energy $\varepsilon$, the electron temperature $T_e$ and  $\omega _e$ are  correlated with each other through
\begin{equation}
\omega_e =n_e\varepsilon=\frac{3}{2}k_B n_e T_e, 
\end{equation}
where $k_B$ is the Boltzmann constant. 
In the energy equation (\ref{eq:EnergyTransport}), the term $-e\bm{\Gamma}_e\cdot\bm{E}$  accounts for the ohmic or joule heating of the electrons in the electric field and the term $S_\omega=S_eH_i$ accounts for the energy loss due to ionization and excitation,  where $H_i$ is the ionization energy.

\section{\label{sec:BCeqs}Boundary Conditions}  

There are three kinds of boundary conditions which are used in plasma modeling: the variable is specified at the boundary (Dirichlet kind), the normal component of the gradient of the variable is specified at the boundary (Neumann kind), or the flux, given by eqs. (\ref{eq:neFlux}-\ref{eq:EnergyFlux}) which is a function involving the variable and its gradient, is specified at the boundary (Robin kind). 
Extrapolation boundary conditions can also be used, although they are typically used for outflows not solid surfaces.

\subsection{\label{sec:PotentialBCs}Electric potential boundary conditions}
The BC for Poisson's equation can be either a Dirichlet kind where the voltage is specified or a Neumann kind where the normal component of the electric field is specified. In this study,  a specified voltage condition is used for the electrodes, which are assumed to be conductors, and a specified electric field condition is used for lateral walls, which are assumed to be dielectrics. 

\subsubsection{\label{sec:grounded}Specified voltage (Dirichlet)}
In the most common configuration for an RF plasma, one electrode has an applied voltage and the other electrode is grounded. Sometimes the outer wall of the reactor will also be grounded \cite{Passchier1993}. For a grounded surface, the boundary condition is:
\begin{equation} 
\phi=0  .
\end{equation}
For the powered electrode, the electric potential  is given by:
\begin{equation} 
\phi=V_\text{DC}+V_\text{RF}\sin\left(2\pi f t\right)  ,
\end{equation}
where $V_\text{DC}$ is the direct current voltage, $V_\text{RF}$ is the radiofrequency voltage and $f$ is the RF frequency. In one-dimensional cases, the direct current voltage will be zero, however, for two or three-dimensional cases, there can be a difference in the area of the powered electrode and the total grounded area, which causes a natural DC bias \cite{Passchier1993,Land2009}. 
In the case of a cylindrical reactor where the outer wall is a dielectric, the DC bias will be zero.

\subsubsection{\label{sec:dielectric}Dielectric (Neumann)}
Another common BC for the electric potential is a dielectric surface \cite{Boeuf1995,Davoudabadi2007}. In this kind, the electric field (gradient of the potential) is imposed through a equation using  the wall charge. The charge density distribution $\sigma$ is the time integral of the current to the wall so for a singly ionized gas, this equation reads:
\begin{equation}\label{eq:dielectric} 
\frac{\partial\sigma}{\partial t}=e\left(\bm{\Gamma}_i-\bm{\Gamma}_e\right)\cdot\bm{n}  ,
\end{equation}
where $\bm{n}$ is the unit normal vector directed out of the domain on the boundary surface. The electric field is correlated with  the wall charge through Gauss' law:
\begin{equation} 
-\bm{E}\cdot\bm{n}=\frac{\sigma}{\epsilon_0}  .
\end{equation}

\subsection{\label{sec:N_BCs}Number and energy density boundary conditions}
The BCs for the number and electric energy equations are either a Dirichlet kind where the number and energy density are specified, or a Robin kind, where the fluxes are specified.  
There is no general consensus in the literature on how to specify the fluxes. Secs. \ref{sec:Drift} through \ref{sec:SecEmis} discusses the flux boundary conditions used in the previous studies. 

\subsubsection{\label{sec:ZeroN}Zero densities (Dirichlet)}
The simplest BC assumes that the surface is perfectly absorbing with no reflection \cite{Passchier1993,Davoudabadi2009}, therefore, the ion and electron number density at the surface are zero, i.e., $n_i=n_e=0$, and correspondingly, $\omega_e=0$. 
In a variant form of these BCs,  the component of ion density gradient normal to the wall is set to zero, i.e., $\bm{\nabla} n_i\cdot\bm{n}=n_e=\omega_e=0$ \cite{Surendra1990}. If the electric field is directed out of the domain causing the ions to flow out of the domain, the zero ion density gradient boundary condition produces the same results as the zero number density boundary condition. This is the case for the plasmas examined in this study so the zero gradient condition is not included here. 
\subsubsection{\label{sec:Drift}Pure drift (Robin)} In this BC, the flux directed towards the surface is assumed pure drift with no diffusion and no flux away from the surface \cite{Davoudabadi2007,Horn2011}.
It is necessary to determine whether the flux is towards the surface, which can be accomplished by defining:
\begin{equation}{\label{eq:aei}}
a_{e(i)}=\left\{ \begin{array}{cc} 1 & \text{sgn}(q)\bm{E}\cdot\bm{n}\geq 0 \\
0 & \text{sgn}(q)\bm{E}\cdot\bm{n}< 0 \end{array} \right. .
\end{equation}
The normal component of the electron (ion) flux at the boundary is:
\begin{equation} \label{eq:DriftBC}
\bm{\Gamma}_{e(i)}\cdot\bm{n}=\left[a_{e(i)}\text{ sgn}(q)\mu_{e(i)}\bm{E}\cdot\bm{n}\right]n_{e(i)},
\end{equation}
where $\bm{\Gamma}_{e(i)}$ is given in eq. (\ref{eq:neFlux}). Correspondingly, the normal component of the energy flux is calculated by:
\begin{equation} \label{eq:DriftBCwe}
\bm{\Gamma}_\omega\cdot\bm{n}=\left[-a_e\frac{5}{3}\mu_e\bm{E}\cdot\bm{n}\right]\omega_e .
\end{equation}

\subsubsection{\label{sec:Thermal}Thermal flux (Robin)} In this BC \cite{Boeuf1995,Paunska2011}, the flux directed towards the surface is a combination of the drift flux to the surface, as in Section \ref{sec:Drift}, and the thermal flux towards the surface. As a result, there is always flux towards the surface even when the electric field causes the drift to be directed away from the surface. The thermal flux towards the surface is assumed to be the one-way flux for a Maxwellian distribution which is equal to $\frac{1}{4}nv_\text{th}$. 
For the ions and electrons, the flux at the boundary is:
\begin{equation} \label{eq:BC3eq}
\bm{\Gamma}_{e(i)}\cdot\bm{n}=\left[a_{e(i)}\text{ sgn}(q)\mu_{e(i)}\bm{E}\cdot\bm{n}+\frac{1}{4}v_{\text{th}_{e(i)}}\right]n_{e(i)} ,
\end{equation}
where $v_{\text{th}_{e(i)}}$ is the thermal velocity of the electrons (ions) determined by:
\begin{equation}
v_{\text{th}_{e(i)}}=\sqrt{\frac{8 k_B}{\pi}\frac{T_{e(i)}}{m_{e(i)}}} .
\end{equation}

\par With respect to the energy BC, there are two  approaches, both of which
 can be formulated by
\begin{equation} \label{eq:we_BC3}
\bm{\Gamma}_\omega\cdot\bm{n}=\left[-a_e\frac{5}{3}\mu_e\bm{E}\cdot\bm{n}+\beta v_{\text{th}_e}\right]\omega_e,
\end{equation}
where $\beta$ is a factor depending on the approach. In the first approach, where the thermal flux is equated to the enthalpy flux \cite{Nitschke1994,Hammond2002,Bogdanov2013}, $\beta =5/12$ as the enthalpy flux is $\frac{5}{2}k_BT_e\bm{\Gamma}_e$. In the second approach, where the thermal flux is equated to the one-way flux of kinetic energy for a Maxwellian distribution \cite{Rafatov2012BC,WenXia2013,Liu2014,Boeuf1995}, $\beta =1/3$ as the one-way flux of kinetic energy for a Maxwellian distribution is $2k_BT_e\bm{\Gamma}_e$. 
The one-way flux of kinetic energy for a Maxwellian distribution is more consistent with the assumptions made for the electron BC, therefore, the second approach is chosen here. 

\subsubsection{\label{sec:Diff}Thermal and diffusion flux (Robin)} 
This boundary condition 
assumes that in addition to the previously considered drift and thermal fluxes, the diffusion flux to the wall is significant. It is based on a formulation proposed by Hagelaar et al.~\cite{Hagelaar2000BC}, but the SEE coefficient is set to zero in order to determine the impact of the diffusion flux. The ion BC adds a diffusion term $\left(-\frac{1}{2}D_i\bm{\nabla}\cdot\bm{n}\right)$ to Eq. (\ref{eq:BC3eq}) giving:
\begin{equation} \label{eq:Dbc}
\bm{\Gamma}_{i}\cdot\bm{n}=\left[a_{i}\mu_{i}\bm{E}\cdot\bm{n}+\frac{1}{4}v_{\text{th}_{i}}\right]n_{i}-\frac{1}{2}D_i\bm{\nabla}n_i\cdot\bm{n} .
\end{equation}
\noindent The term involving the gradient can be challenging to implement due to possible numerical difficulties in evaluating the gradient, so 
Hagelaar et al.~\cite{Hagelaar2000BC} proposed an alternative form, using the definition of the ion flux in Eq.~(\ref{eq:neFlux}). 
This form for the electron (ion) flux reads:
\begin{equation} \label{eq:ni_BC45}
\bm{\Gamma}_{e(i)}\cdot\bm{n}=\left[\text{sgn}(q)\left(2a_{e(i)}-1\right)\mu_{e(i)}\bm{E}\cdot\bm{n}+\frac{1}{2}v_\text{th}\right]n_{e(i)} .
\end{equation}
\noindent Since the plasma model used by Hagelaar et al.~\cite{Hagelaar2000BC} did not include the electron energy equation, they did not discuss the BC for the energy equation. Here, the following condition, which is consistent with Eq.~(\ref{eq:Dbc}),  is used for the energy flux of the electrons:

\begin{equation} 
\bm{\Gamma}_{\omega}\cdot\bm{n}=\left[-a_e\frac{5}{3}\mu_e\bm{E}\cdot\bm{n}+\frac{1}{3} v_{\text{th}_e}\right]\omega_e 
-\frac{5}{6}D_e\bm{\nabla}\omega_e\cdot\bm{n} .
\end{equation}

\noindent
The last term accounts for the diffusion, the coefficient is $D_\omega/2$ which simplifies to $5D_E/6$ since $D_\omega=5D_e/3$, the other terms match Eq.~(\ref{eq:we_BC3}) for the thermal flux boundary condition. 
Another equivalent alternative form without the gradient term is given by: 
\begin{equation} \begin{gathered} \label{eq:we_BC4}
\bm{\Gamma}_{\omega}\cdot\bm{n}=\left[-\left(2a_e-1\right)\frac{5}{3}\mu_{e}\bm{E}\cdot\bm{n}+\frac{2}{3}v_\text{th}\right]\omega_{e}  .
\end{gathered} \end{equation} 

\subsubsection{\label{sec:SecEmis}Secondary electron emission (Robin)} This BC includes the effect of the SEE by ion impact. Here, the formulation given in  the previous works~\cite{Hagelaar2000BC,Lee2012,Iqbal2014} are used to apply this BC. 
The ion flux at the boundary is identical to Eq.~(\ref{eq:ni_BC45}).
On the other hand, the net electron density at the boundary is the combination of the density of SEE electrons, $n_\gamma$, and the density of primary electrons directed from the bulk, $n_\alpha=n_e-n_\gamma$. Therefore, the electron flux normal to the wall is given by $\bm{\Gamma}_{e}\cdot\bm{n}=\bm{\Gamma}_\gamma\cdot\bm{n}+\bm{\Gamma}_{\alpha}\cdot\bm{n}$, where $\bm{\Gamma}_\gamma$ is the flux of the SEE electrons and $\bm{\Gamma}_{\alpha}$ is the bulk electron flux. The BC for the bulk electron flux is similar to that for the ions:
\begin{equation} \begin{gathered} \label{eq:ne_bulk}
\bm{\Gamma}_{\alpha}\cdot\bm{n}=\left[-\left(2a_e-1\right)\mu_{e}\bm{E}\cdot\bm{n}+\frac{1}{2}v_\text{th}\right]n_{\alpha} .
\end{gathered} \end{equation}

\noindent
On the other hand, the secondary electrons are assumed to have a beam-like behavior and not flow back to the wall, thus, the BC for the SEE flux is:
\begin{equation}\label{eq:SEEFlux}
\bm{\Gamma}_\gamma\cdot\bm{n}=-\left(1-a_e\right)\gamma\bm{\Gamma}_{i}\cdot\bm{n}  , 
\end{equation}
where $\gamma$ is the SEE coefficient which defines the average number of electrons emitted per ion impact.
Due to the beam-like behavior assumption for the emitted secondary electrons, they do not flow back towards the surface and diffusion can be neglected. Therefore, the number density of the secondary electrons can be written as:
\begin{equation} \label{eq:ne_emit}
n_\gamma=\left(1-a_e\right)\frac{\gamma\bm{\Gamma}_{i}\cdot\bm{n}}{\mu_e\bm{E}\cdot\bm{n}}.
\end{equation}
From the equations above, the BC for the total electron flux is derived:
\begin{equation} \begin{gathered} \label{eq:neBC5}
\bm{\Gamma}_{e}\cdot\bm{n}=\left[-\left(2a_e-1\right)\mu_{e}\bm{E}\cdot\bm{n}+\frac{1}{2}v_\text{th}\right]n_{e}
\\ -\frac{1}{2}v_\text{th}n_\gamma-2\left(1-a_e\right)\gamma\bm{\Gamma}_i\cdot\bm{n}.
\end{gathered} \end{equation}
The detail of this derivation is given in  Appendix \ref{sec:seeBCDerive}. Eq.~(\ref{eq:neBC5}) is identical to the SEE BC equation used in the previous works ~\cite{Hagelaar2000BC,Lee2012,Iqbal2014}. 

The energy density at the boundary consists of the SEE electron energy $\omega_\gamma$ and the bulk energy $\omega_\alpha=\omega_e-\omega_\gamma$. The energy flux of the electrons from the bulk is calculated by: 
\begin{equation} \begin{gathered} \label{eq:we_bulk}
\bm{\Gamma}_A\cdot\bm{n}=\left[-\left(2a_e-1\right)\left(\frac{5}{3}\mu_{e}\right)\bm{E}\cdot\bm{n}+\frac{2}{3}v_\text{th}\right]\omega_{\alpha}  .
\end{gathered} \end{equation} 
Due to the beam-like behavior assumed for the SEE electrons, the energy BC for the SEE electrons is:
\begin{equation}
\bm{\Gamma}_B\cdot\bm{n}=\left[-\left(1-a_e\right)\gamma\bm{\Gamma}_{i}\cdot\bm{n}\right]\varepsilon_\gamma  , 
\end{equation}
where $\varepsilon_\gamma$ is the mean energy that the secondary electrons are emitted at, which for this study is set to 2 eV \cite{Becker2013}. 
The energy density of the secondary electrons is calculated by:
\begin{equation} \label{eq:we_emit}
\omega_\gamma=\left(1-a_e\right)\frac{\gamma\varepsilon_\gamma\bm{\Gamma}_{i}\cdot\bm{n}}{\mu_e\bm{E}\cdot\bm{n}} .
\end{equation}
Using Eqs (\ref{eq:we_bulk}-\ref{eq:we_emit}) and following a derivation procedure similar to the one given for the electron BC in Eq.~(\ref{eq:neBC5}), the following equation is derived for the energy BC:
\begin{equation} \begin{gathered} \label{eq:we_BC5}
\bm{\Gamma}_{\omega}\cdot\bm{n}=\left[-\left(2a_e-1\right)\frac{5}{3}\mu_e\bm{E}\cdot\bm{n}+\frac{2}{3}v_\text{th}\right]\omega_{e}
\\ -\frac{2}{3}v_\text{th}\omega_\gamma-2\left(1-a_e\right)\varepsilon_\gamma\gamma\bm{\Gamma}_i\cdot\bm{n}.
\end{gathered} \end{equation}

\noindent
This equation is identical to that used by Lee et al.~\cite{Lee2012} to set the energy BC. 

\subsection{\label{sec:IonBC} Extrapolation boundary condition for ions}
In the boundary conditions given in Sections \ref{sec:ZeroN}-\ref{sec:SecEmis}, the ions and electrons are dealt with at the boundary, similarly. However, Hammond \textit{et al} \cite{Hammond2002} suggested that this treatment may not be appropriate because their analysis showed that ion boundary conditions were not necessary. Their analysis showed that the electric field was directed out of the plasma which meant the ions always flowed out of the domain. To set up a boundary condition consistent with the suggestion made by Hammond \textit{et al}, cases were run in the current study with an extrapolation condition used for the ion number density when the electric field was directed out of the domain. In the extrapolation boundary condition, the value of the variable at the boundary is extrapolated from the values at the grid points near the boundary.

\section{\label{sec:NumMethod}Numerical Method}  

\par A second order discretization scheme was applied on the time derivatives in the plasma equations:

\begin{equation}
\frac{3n_e^{n+1}-4n_e^n+n_e^{n-1}}{2\Delta t} 
+\bm{\nabla}\cdot\left(-\mu_en_e^{n+1}\bm{E}^{n+1}-D_e\bm{\nabla}n_e^{n+1}\right)=S_e^{n} ,
\label{eq:neDisc}
\end{equation}
\begin{equation}
\frac{3n_i^{n+1}-4n_i^n+n_i^{n-1}}{2\Delta t} 
+\bm{\nabla}\cdot\left(\mu_in_i^{n+1}\bm{E}^{n+1}-D_i\bm{\nabla}n_i^{n+1}\right)=S_e^{n} ,
\label{eq:niDisc}
\end{equation}
\begin{equation} \begin{gathered}
\frac{3\omega_e^{n+1}-4\omega_e^n+\omega_e^{n-1}}{2\Delta t} 
+\bm{\nabla}\cdot\frac{5}{3}\left(-\mu_e\omega_e^{n+1}\bm{E}^{n+1}-D_e\bm{\nabla}\omega_e^{n+1}\right)
\\ =-e\bm{\Gamma}_e^{n+1}\cdot\bm{E}^{n+1}+S_\omega^{n} ,
\end{gathered} 
\label{eq:energyDisc}
\end{equation} 
\begin{equation}
\nabla^2\phi^{n+1}=\frac{e}{\epsilon_0}\left( n_e^{n+1}-n_i^{n+1}\right),
\label{eq:ptentialDisc}
\end{equation}
where the superscript $n$ indicates the time level and the electric field comes from Eq. (\ref{eq:E_inst}). The ionization source is treated as an explicit term.
These equations constitute a nonlinear set of equations at time level $n+1$ for $n_e^{n+1}$,  $n_i^{n+1}$, $\omega_e^{n+1}$, and $\phi^{n+1}$ after the spatial discretization is carried out. The nonlinearity is due to the nonlinear first terms in the parentheses of eqs.~(\ref{eq:neDisc}-\ref{eq:energyDisc}) on the left hand sides and the nonlinear first term on the right hand side of eq.~(\ref{eq:energyDisc}). The equations are advanced in time by estimating the electric field  at the new timestep, $\bm{E}^{n+1}$, then updating $n_e^{n+1}$,  $n_i^{n+1}$, $\omega_e^{n+1}$ and finally the electric field is updated. Since $n_e$ is updated before $\omega_e$, the ohmic heating term in Eq. (\ref{eq:energyDisc}) uses the updated value of $n_e$. The equations are iterated until the system is solved simultaneously.
\par For the spatial discretization, a finite difference method is applied in an axisymmetric cylindrical coordinate consistent with the axisymmetric  geometry of the plasma reactor studied in this work. 
The spatial discretization of the fluxes, the terms in the parentheses of eqs (\ref{eq:neDisc}-\ref{eq:ptentialDisc}), uses the Scharfetter-Gummel scheme~\cite{Scharfetter1969,Boeuf1987,Hagelaar2000}. 
For this scheme, the equation for the flux of $n$ in the y-direction is given by:
\begin{equation} \begin{gathered} \label{eq:SGfluxEq}
\Gamma_{j+1/2}=\frac{D}{\Delta y} \left[\frac{\text{Pe}}{1-\exp\left(-\text{Pe}\right)}n_j-\left(\frac{\text{Pe}} {1-\exp\left(-\text{Pe}\right)}-\text{Pe}\right) n_{j+1}\right] ,
\end{gathered} \end{equation}
where $\text{Pe}$ is the Peclet number, defined as:
\begin{equation}
\text{Pe}=\frac{\text{sgn}(q)\mu E_y \Delta y}{D} .
\end{equation}
In  the computations with very small values of $\mathrm{Pe}$, an expanded expression is used for $ \frac{\mathrm{Pe}}{1-\exp\left(-\mathrm{Pe}\right)}=1+\mathrm{Pe}/{2}+{\mathrm{Pe}^2}/{12}-{\mathrm{Pe}^4}/{720}+\cdots
$.
\par
\begin{figure}[h]
  \centering
    \includegraphics[scale=0.3]{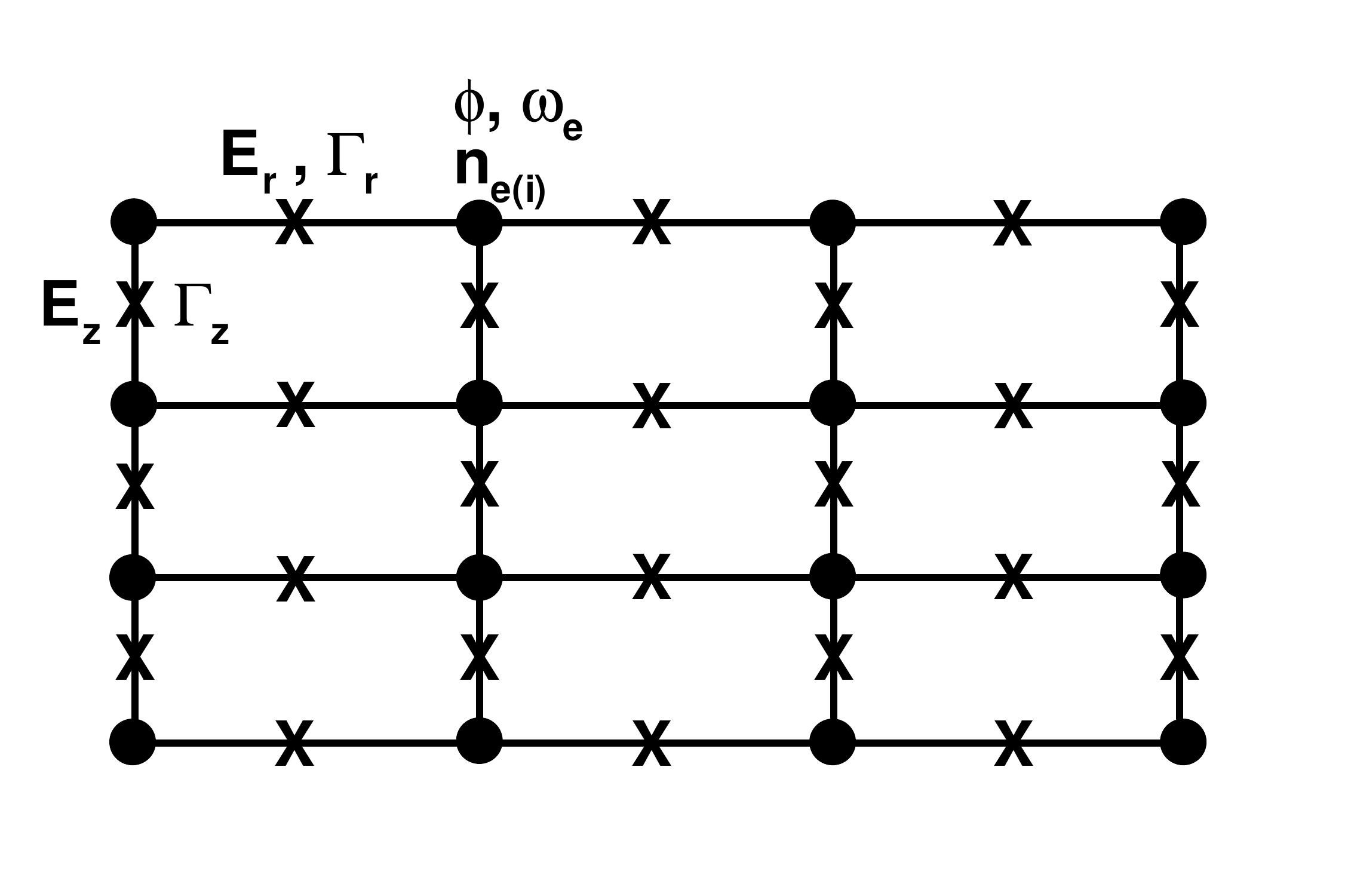}
    \caption{Staggered finite difference mesh: primary variables are stored at nodes (designated by $\bullet$) and fluxes and the electric field are stored at the midpoints (designated by $\times$). }
    \label{fig:grid}
\end{figure}

 A nonuniform, tan-stretched, grid is used with a higher resolution near the electrodes and outer wall. The grid is staggered, as shown in Fig. \ref{fig:grid}, with the primary variables, $\phi$, $n_{e,i}$, and $\omega_e$, evaluated on the nodes, j, and the electric field $\bm{E}$ and the fluxes  $\bm{\Gamma}_{e,i,\omega}$ evaluated halfway between the nodes. The boundary passes through the nodes. For the boundary conditions where the flux to the surface is considered, the flux is specified at the midpoint between the boundary and the first interior node and the number and energy density at the boundary is obtained from Eq. (\ref{eq:SGfluxEq}). 

\section{\label{sec:Results}Results and Discussion}  

\subsection{Validation of plasma model}
\par
\begin{figure}[h]   
  \centering
  \begin{subfigure}[t]{0.45\textwidth}
    \centering
    \includegraphics[scale=0.30]{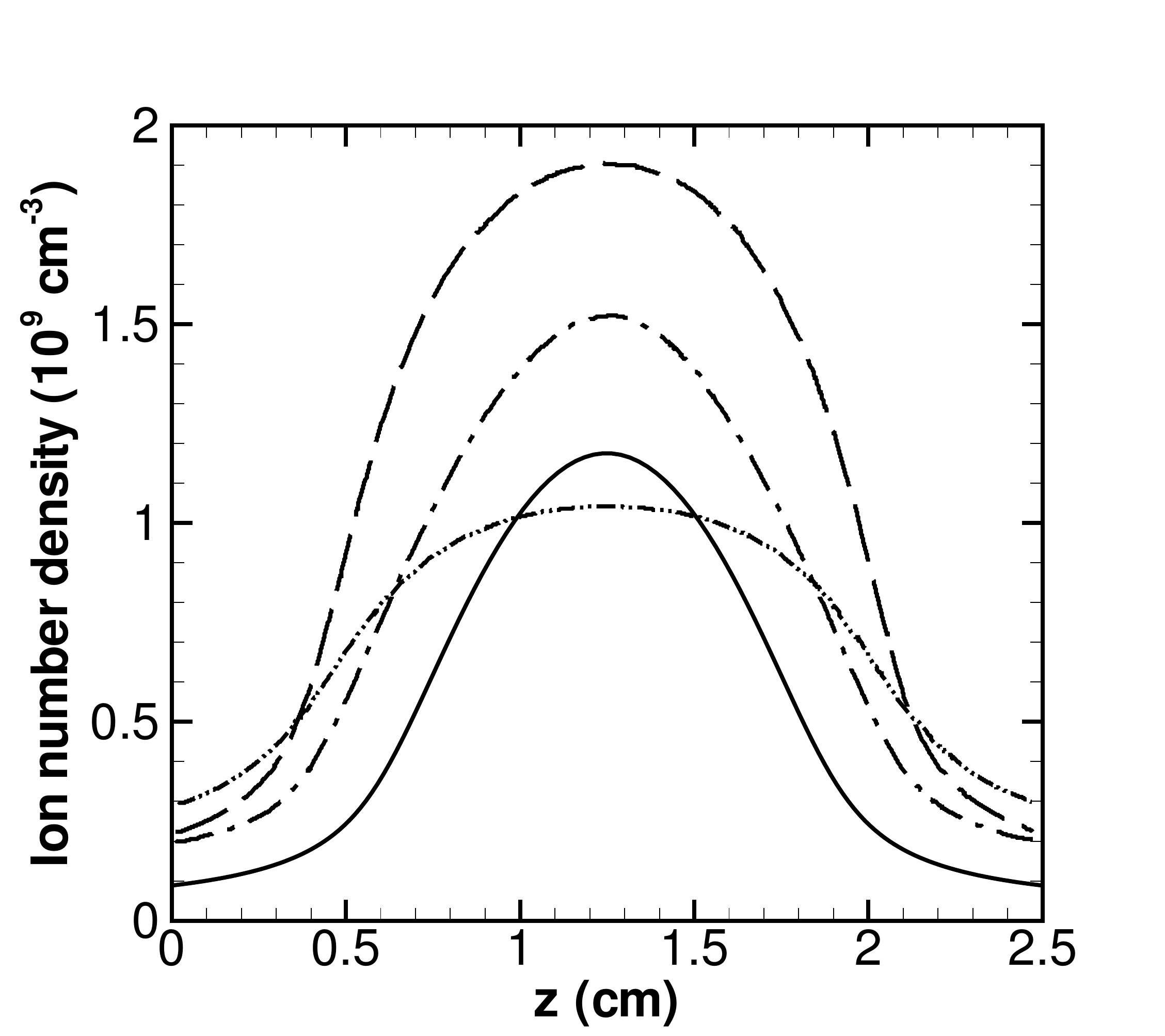}
    \caption{}
    \label{fig:PIC150}
  \end{subfigure}
 ~
  \begin{subfigure}[t]{0.45\textwidth}   
    \centering
    \includegraphics[scale=0.30]{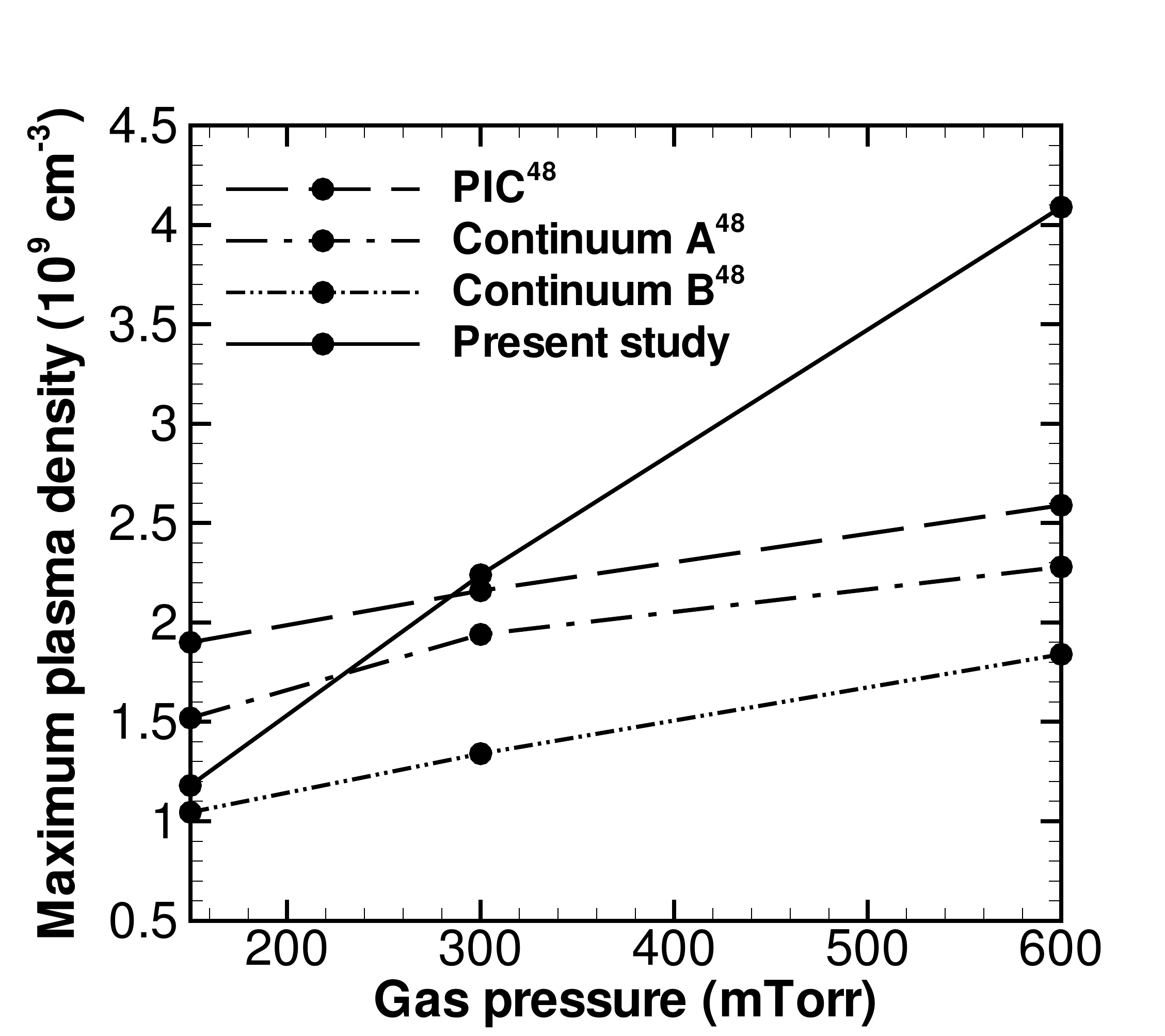}
    \caption{}
    \label{fig:PICvsP}
  \end{subfigure}
  \caption{Validation of the continuum model used in the current study against the PIC and two continuum models A (DDAn) \& B (DDA53) of Becker et al.\cite{Becker2017} for an argon plasma: (a) RF-averaged ion number density (in 10$^9$ cm$^{-3}$) for a gas pressure of 150 mTorr; and (b) maximum plasma density (in 10$^9$ cm$^{-3}$) as a function of gas pressure (in mTorr).}
  \label{fig:PICcomparison}
\end{figure}

 The plasma model was validated against the results of Becker et al. \cite{Becker2017}, who examined the differences between PIC/MCC and two continuum model simulations in a one dimensional reactor setup. The continuum model used here differs from their continuum models in the assumption of constant transport coefficients and the source of the ionization rate as well as the modeling of the ion flux. The models were compared at three different gas pressures, 150, 300, and 600 mTorr. Becker et al. specified the amplitude of the electrode current density as 10 $A/m^2$ which, for the PIC/MCC simulations, corresponded to voltage amplitudes of 90, 70, and 60V for the three pressures, respectively. Their continuum models produced different voltages as they matched the amplitudes of the current density. Here, we matched the voltage from their PIC/MCC simulations. Figure \ref{fig:PIC150} shows the ion number density for a gas pressure of 150 mTorr and Fig. \ref{fig:PICvsP} shows the maximum plasma density as a function of gas pressure. The current model underpredicts the maximum plasma density by 40\% compared to the PIC/MCC code at the lowest pressure but agrees well for the 300 mTorr case; the 60\% difference for the 600 mTorr case is likely due to matching the voltage rather than the current. Since specifying the current to the electrodes, rather than the voltage, adds an additional complication to the boundary condition comparison, this study specifies the electrode voltage and does not attempt to compare the effect of the boundary conditions with PIC/MCC simulations.

To examine the impact of the boundary condition  on the solution, an argon plasma was modeled using each of the number and energy density boundary conditions discussed in Section \ref{sec:N_BCs}. Two configurations of the plasma reactor were considered: a one-dimensional setup and an axisymmetric cylindrical one.  Both the one-dimensional and cylindrical configurations had the RF voltage applied to the lower electrode while the upper electrode was grounded.  The wall in the cylindrical configuration was dielectric meaning that the DC bias voltage discussed in Section \ref{sec:PotentialBCs} is zero. The operating parameters are tabulated in Table \ref{tab:parameters}. The mobility and diffusion coefficients were assumed constant. 

\par Several cases were simulated with different sets of the BCs, as illustrated in Secs.~\ref{sec:ZeroN}\textendash\ref{sec:SecEmis}. For ease of reference, the number and energy boundary conditions are designated as shown in Table \ref{tab:BClabels}. 
All cases were simulated until a quasi-steady state was reached which is defined as the sate when the change in the RF-period averaged values was negligible. Here, BC4 is treated as the baseline case to which the other boundary conditions are compared.

\subsection{One-dimensional configuration results}
\begin{figure}[h]   
  \centering
  \begin{subfigure}[t]{0.45\textwidth}
    \centering
    \includegraphics[scale=0.25]{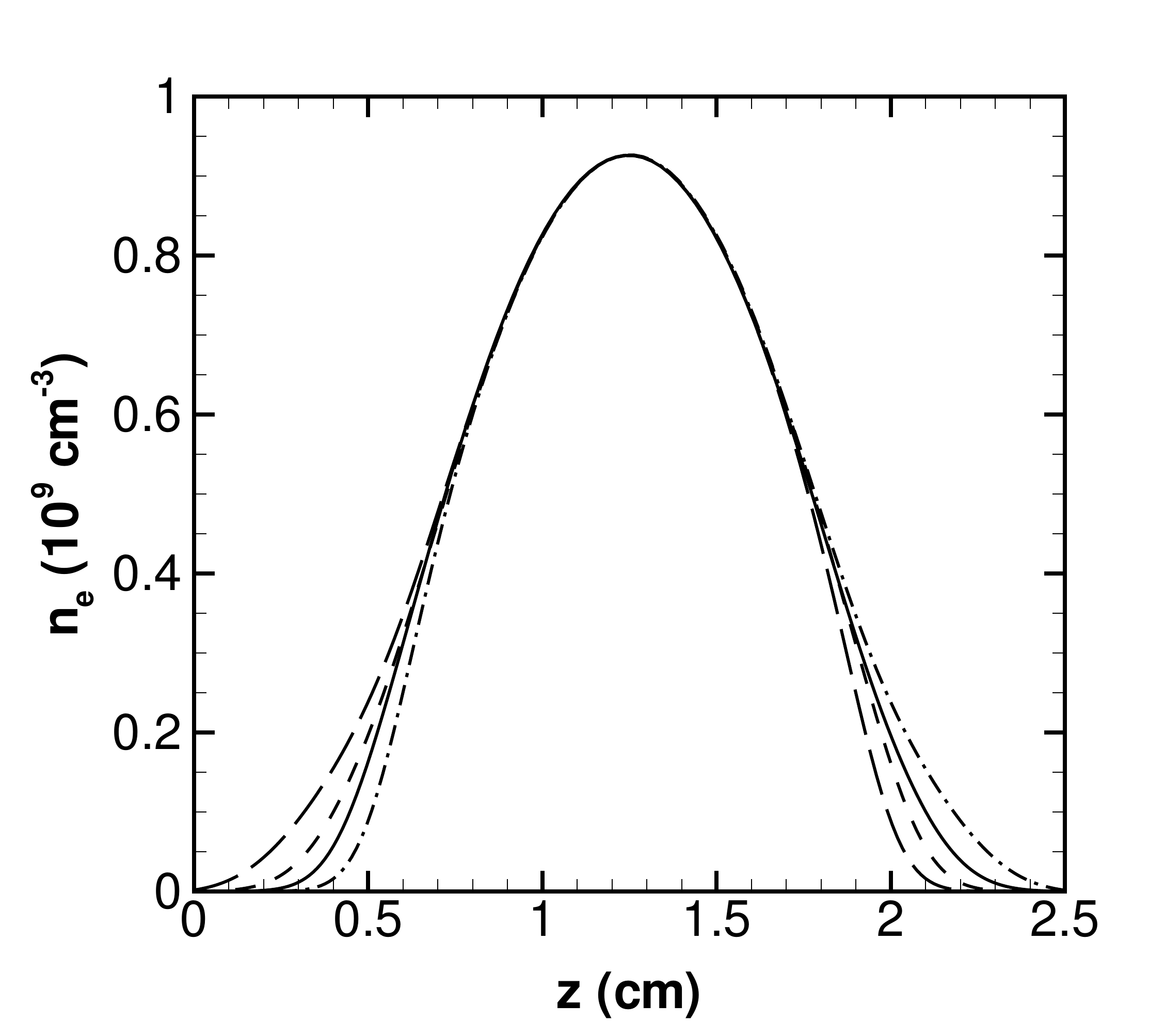}
    \caption{}
    \label{fig:ne_time}
  \end{subfigure}
 ~
  \begin{subfigure}[t]{0.45\textwidth}   
    \centering
    \includegraphics[scale=0.25]{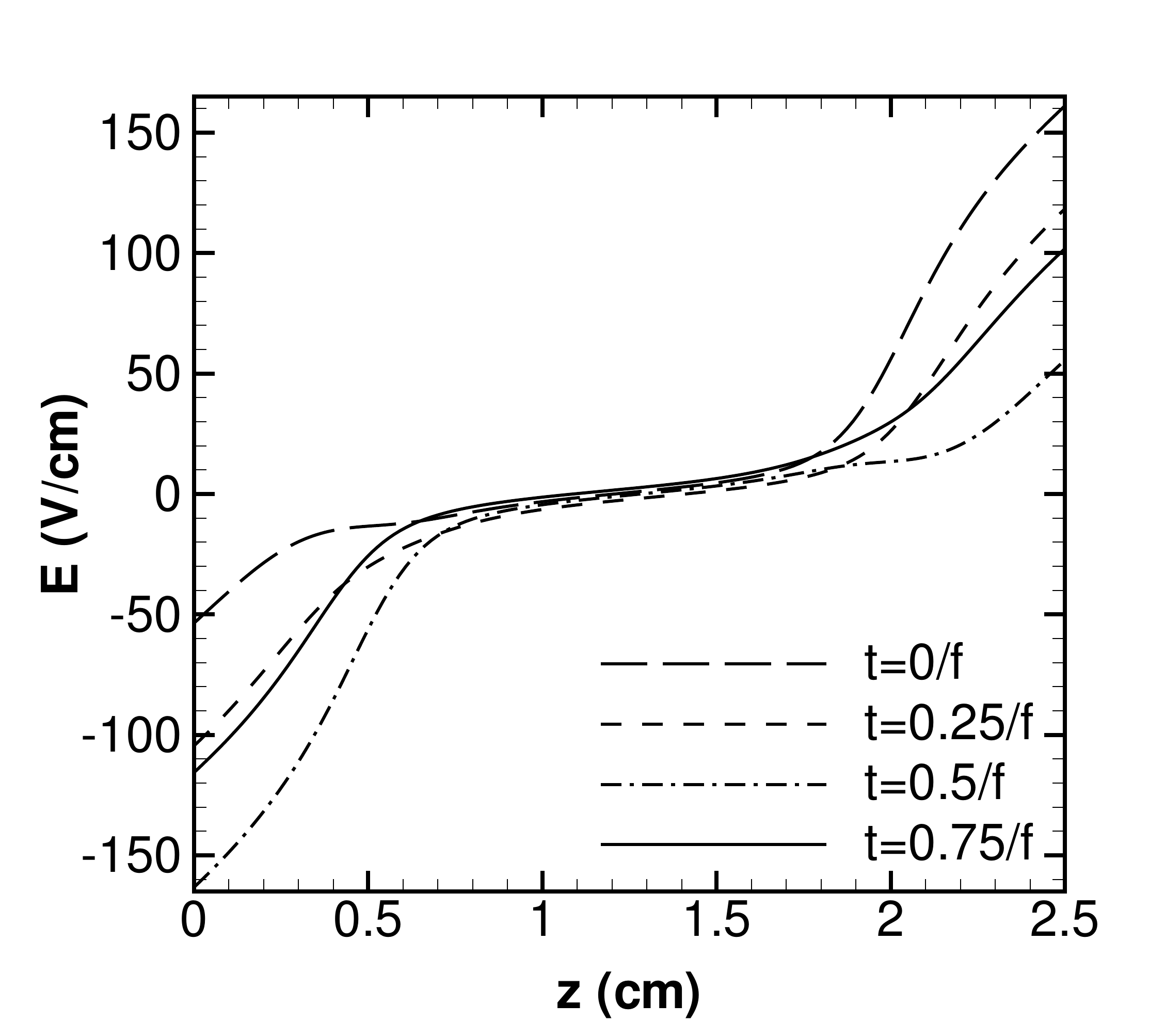}
    \caption{}
    \label{fig:Ey_time}
  \end{subfigure}
  \caption{Spatial variation of (a) electron number density (in 10$^9$ cm$^{-3}$); and (b) Electric field magnitude (in V/cm) at different times in the one-dimensional configuration}
  \label{fig:1Dtime}
\end{figure}

 Figure~\ref{fig:1Dtime} shows the spatial variation of the instantaneous electron number density and electric field at four different times during the RF period for BC4. All other cases exhibit qualitatively similar trends so the transient results are only shown for the BC4 case. 
Both electrodes have large sheaths, the time variation of the electron number density is significant in the region less than 0.75 cm from the electrode.  
The electric field  is negative at the lower electrode, varying between -50  and -150 V/cm, and positive at the upper electrode, varying between 50 and 150 V/cm. This behavior indicates that the electric field is directed out of the domain through the entire RF cycle for both electrodes.

\begin{figure}[h]   
  \centering
  \begin{subfigure}[t]{0.45\textwidth}
    \centering
    \includegraphics[scale=0.24]{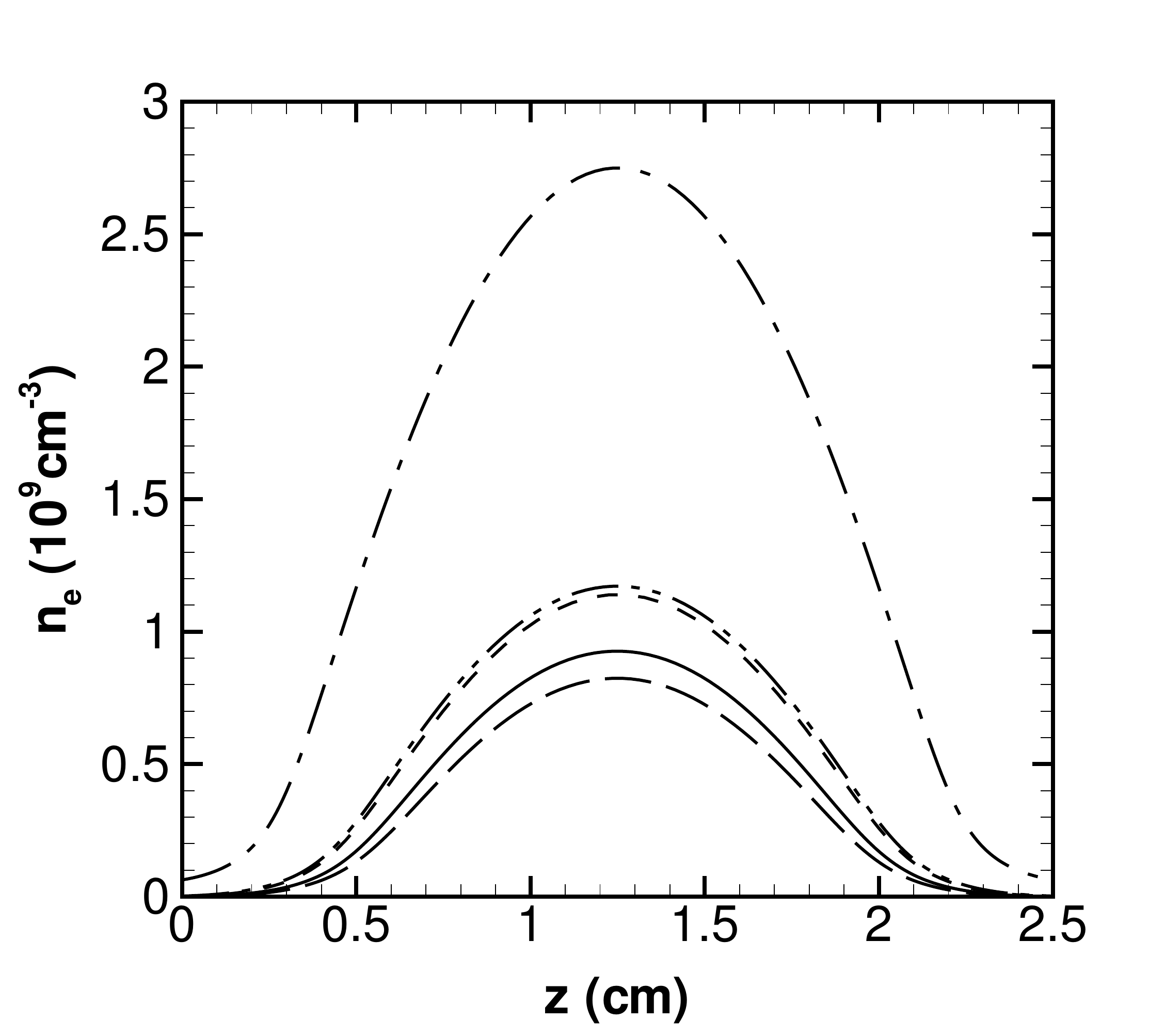}
    \caption{}
    \label{fig:ne_1D}
  \end{subfigure}
  ~
  \begin{subfigure}[t]{0.45\textwidth}   
    \centering
    \includegraphics[scale=0.24]{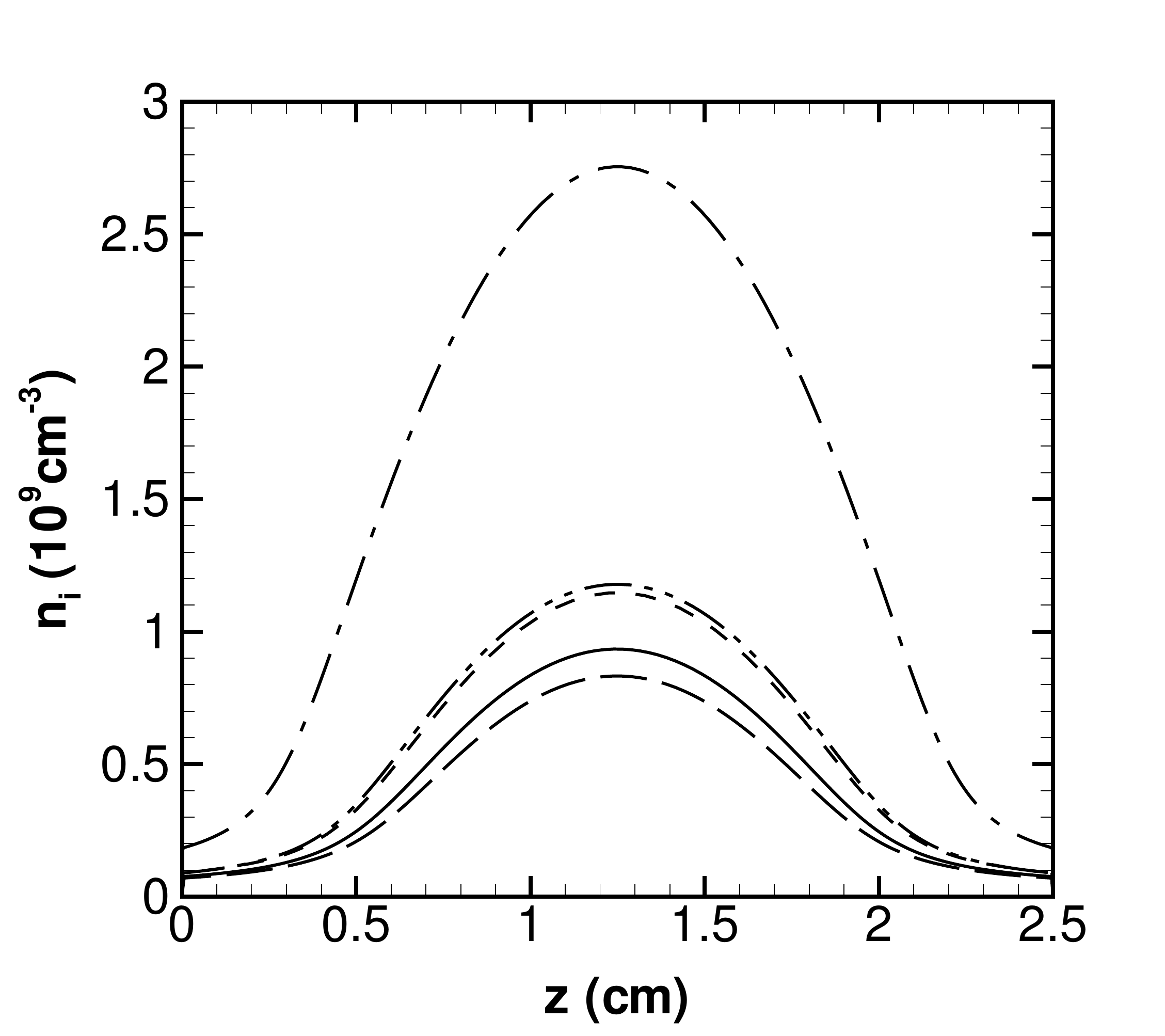}
    \caption{}
    \label{fig:ni_1D}
  \end{subfigure}
  
  \begin{subfigure}[h]{0.45\textwidth}    
    \centering
    \includegraphics[scale=0.24]{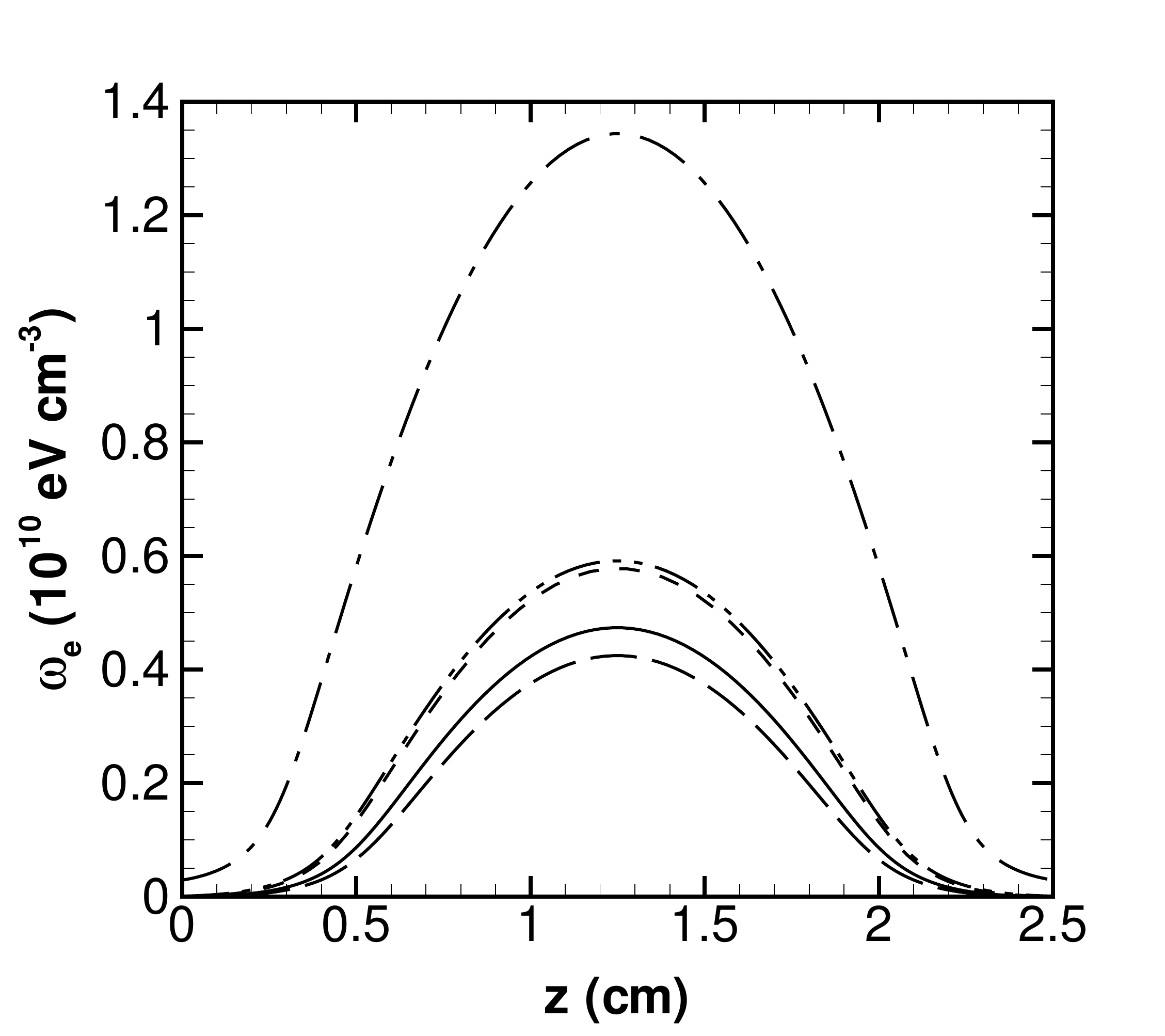}
    \caption{}
    \label{fig:we_1D}
  \end{subfigure}
  ~
  \begin{subfigure}[h]{0.45\textwidth} 
    \centering
    \includegraphics[scale=0.24]{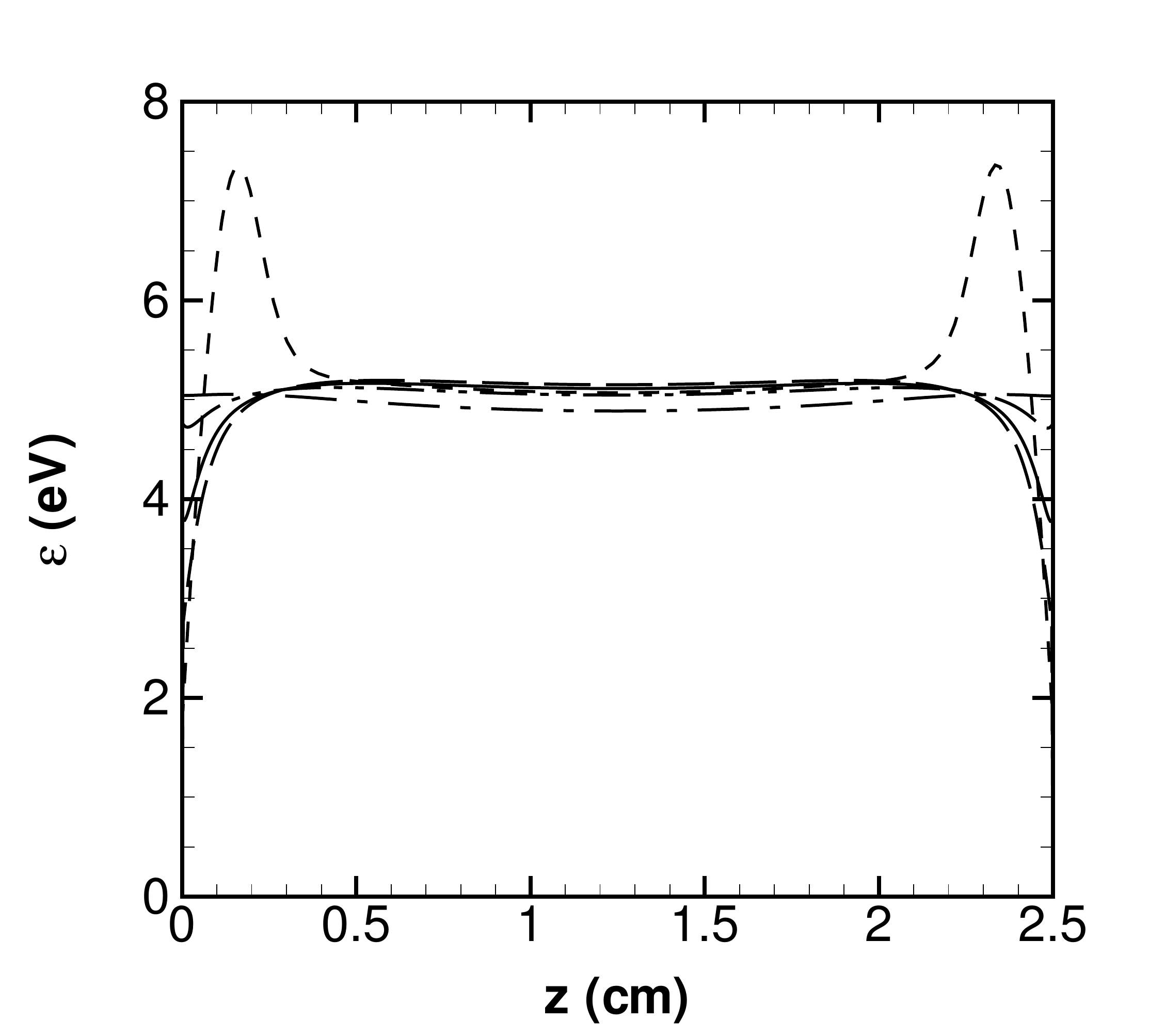}
    \caption{}
    \label{fig:eps_1D}
  \end{subfigure}
  
  \begin{subfigure}[h]{0.45\textwidth}
    \centering
    \includegraphics[scale=0.24]{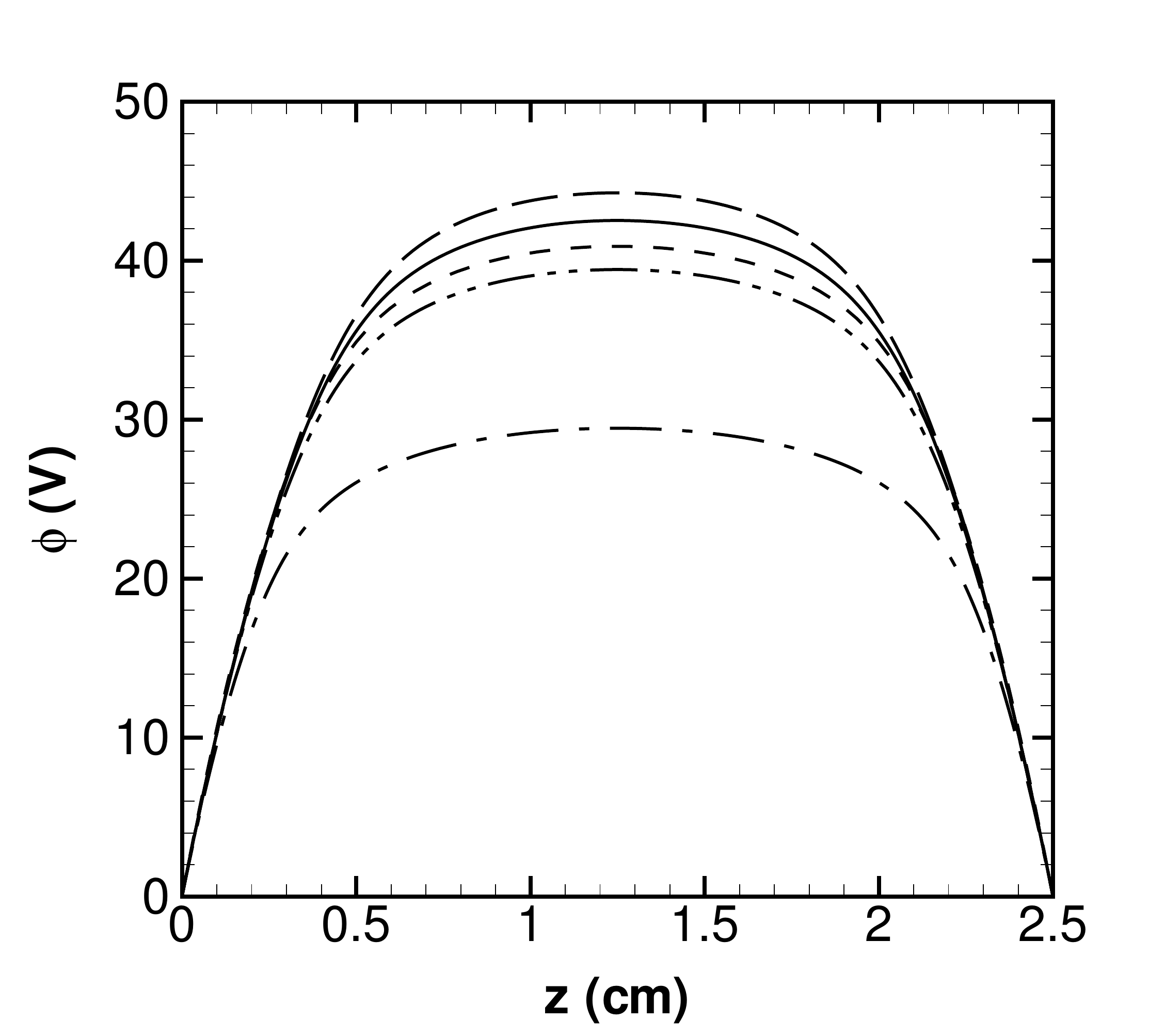}
    \caption{}
    \label{fig:potential_1D}
  \end{subfigure}
  ~
  \begin{subfigure}[h]{0.45\textwidth}  
    \centering
    \includegraphics[scale=0.24]{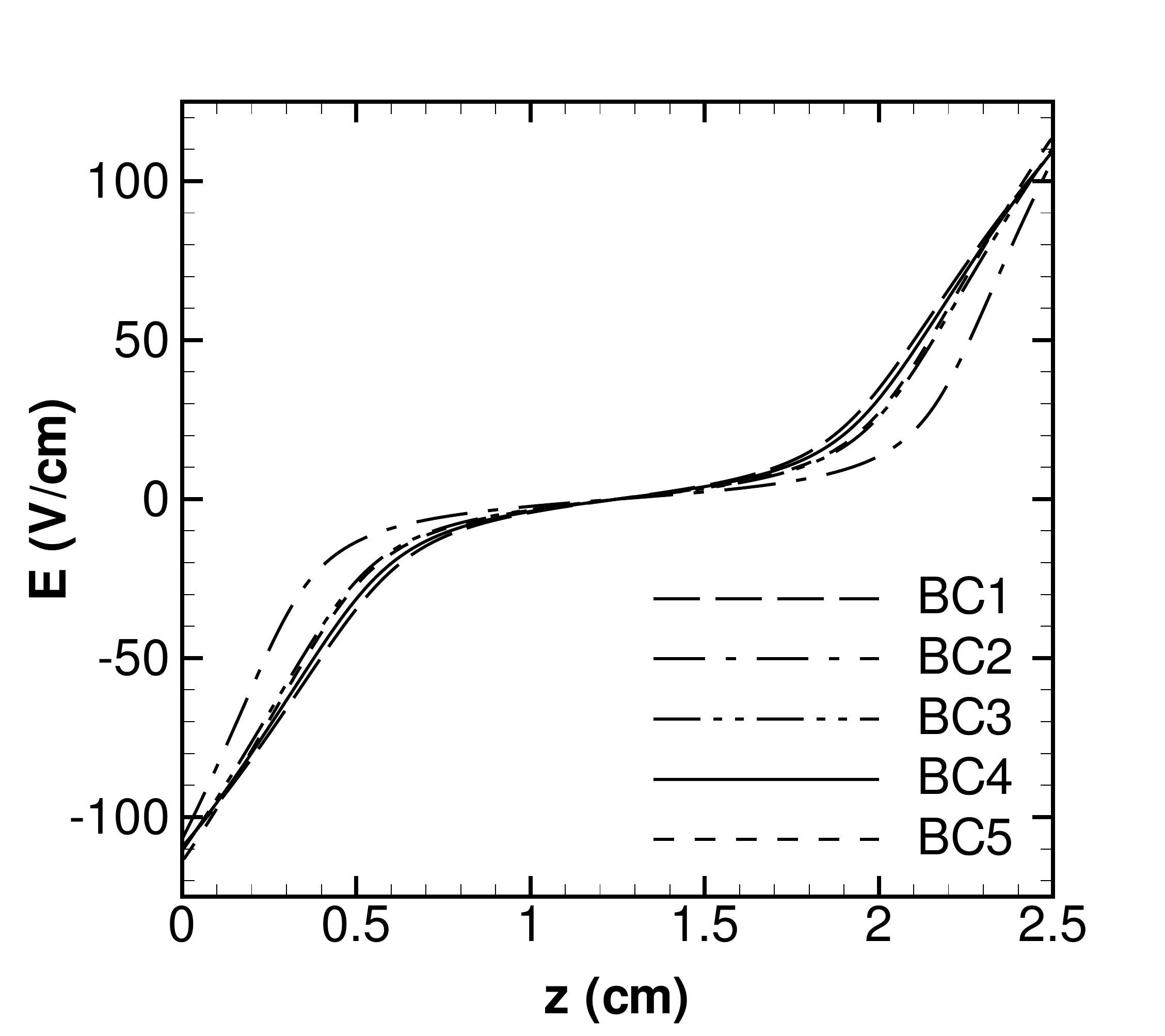}
    \caption{}
    \label{fig:Ey_1D}
  \end{subfigure}
  \caption{Spatial variation of RF-averaged plasma variables: (a) electron number density (in 10$^9$ cm$^{-3}$); (b) ion number density (in 10$^9$ cm$^{-3}$); (c) electron energy density (in 10$^{10}$ eV cm$^{-3}$); (d) mean electron energy (in eV); (e) electric potential (in V); and (f) electric field magnitude (in V/cm) in the one-dimensional configuration.}
  \label{fig:1Dresults}
\end{figure}

Figure~\ref{fig:1Dresults} shows the RF-averaged plasma variables for the different BC cases. 
As seen in Figs.~\ref{fig:ne_1D} and  \ref{fig:ni_1D},  BC2 (the pure drift BC) has a substantially higher plasma density than the other BC kinds. 
This is due to the fact that the electric field is always pointing out of the domain so the electron BC uses a zero flux condition which results in a higher electron density at the boundary because the surface does not act as an electron sink. In a practical situation, flux towards the surface is always expected due to thermal motion and diffusion. It is seen that BC4 produces a lower plasma density than BC3, which is  due to the lacking diffusion flux in BC3. 
The difference between BC4 and BC5 results  in Fig.~\ref{fig:1Dresults} is significant, showing a  significant influence of the SEE mechanism for the considered SEE coefficient of 0.5.
All cases produce a bulk mean electron energy of about 5eV. The case with SEE, BC5, shows a substantial increase in the mean energy in part of the sheath, which has been observed in other studies \cite{Lymberopoulos1993} as well. 
\par In Section \ref{sec:IonBC}, it was noted that there were potential concerns about the ion boundary condition due to the electric field causing the ions to flow out of the domain. Fig.~\ref{fig:Ey_1D} shows that all BC's have the RF-averaged electric field directed out of the domain and Fig.~\ref{fig:Ey_time} shows that the electric field is directed out of the domain throughout the  RF period so both electrodes meet the criteria for setting the ion extrapolation boundary condition. However, when the ion BC was changed to the extrapolation condition, none of the cases experienced an appreciable impact in the bulk of the plasma. This change in the ion boundary condition only affected the ion number density on the boundary itself, the interior points were unaffected. 
\par
\begin{figure}[h]   
  \centering
  \begin{subfigure}[t]{0.45\textwidth}
    \centering
    \includegraphics[scale=0.24]{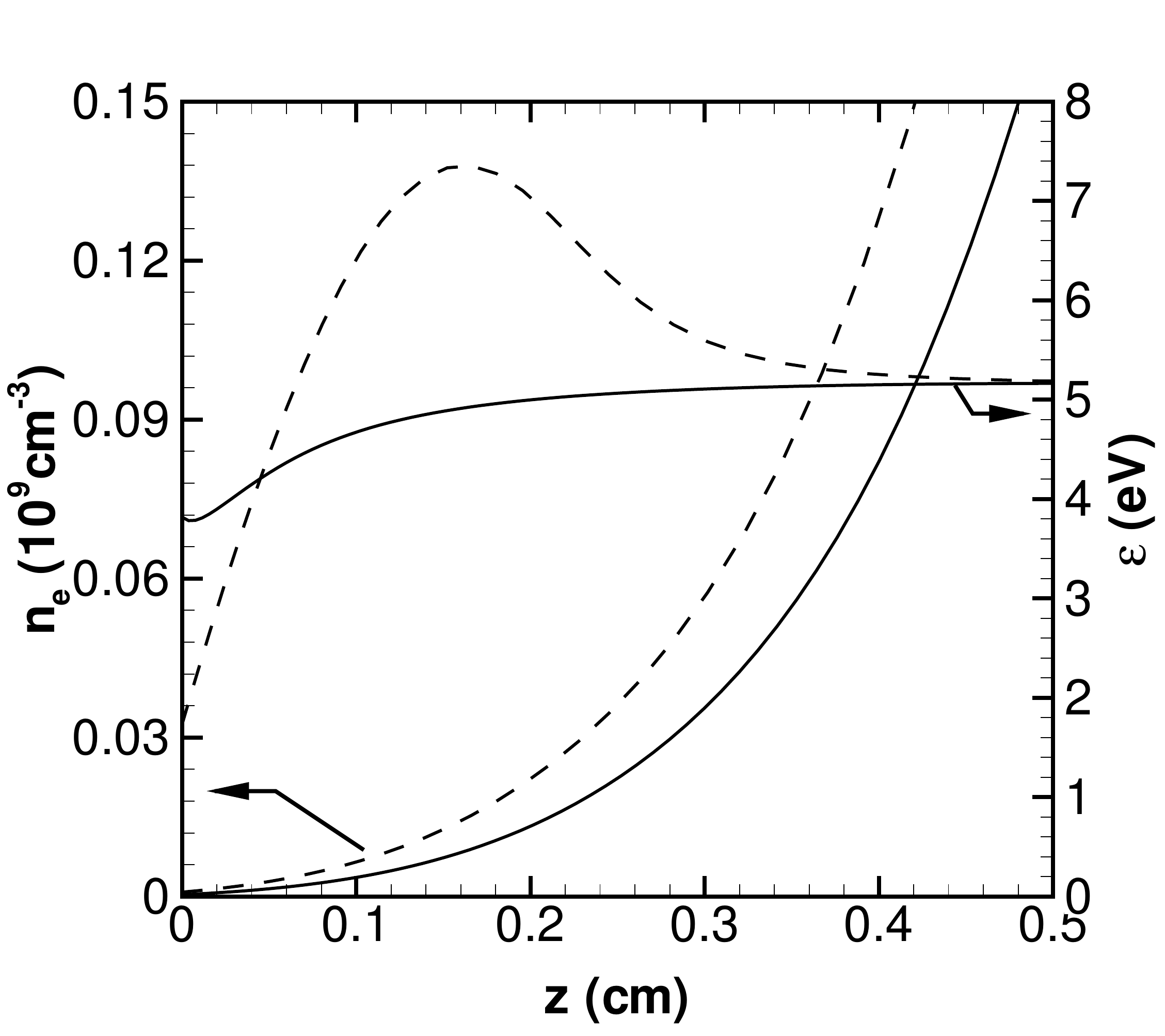}
    \caption{}
    \label{fig:BC5_sheath1}
  \end{subfigure}
  ~
  \begin{subfigure}[t]{0.45\textwidth}   
    \centering
    \includegraphics[scale=0.24]{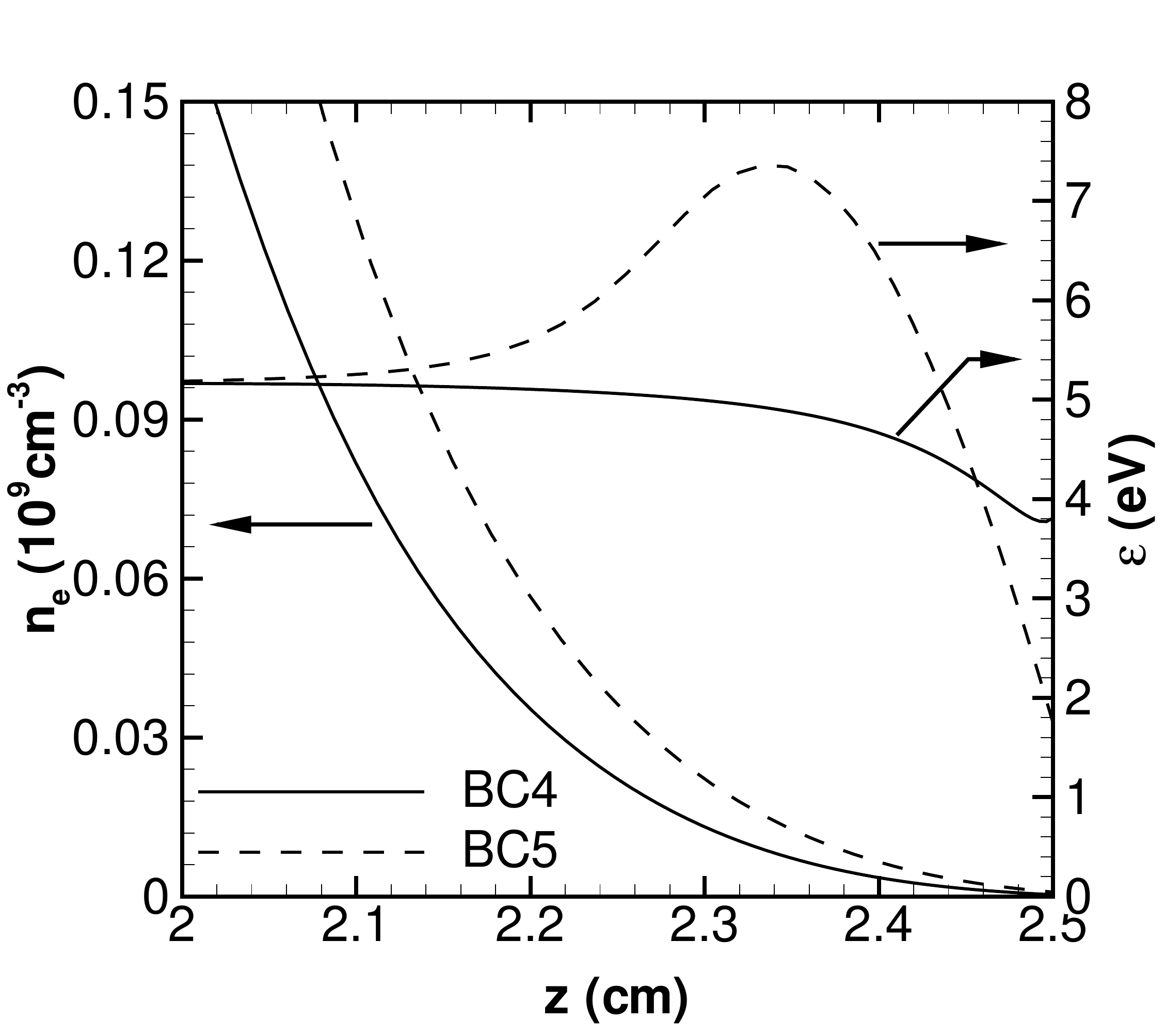}
    \caption{}
    \label{fig:BC5_sheath2}
  \end{subfigure}
  \caption{Spatial variation of electron number density (in 10$^9$ cm$^{-3}$); and mean energy (in eV) for the (a) lower sheath and (b) upper sheath.}
  \label{fig:BC5_sheath}
\end{figure}

 The SEE coefficient $\gamma$ in BC5 was first set to 0.05, which is a typical value used in DC glow discharges \cite{Hagelaar2000BC}. With this value, the results were less than 1\% different from the BC4 case, which is different from BC5 only in neglecting the SEE. Therefore, the SEE coefficient was increased by a factor of ten to determine an upper bound on the possible impact.
Setting $\gamma=0.5$ resulted in a 23\% increase in the maximum electron number density in the bulk. This increase is attributed to increased ionization in the sheath, which in turn is due to a substantially higher mean energy in that region when SEE effects are included (compare BC4 and BC5 in Fig.~\ref{fig:BC5_sheath}).
Since the secondary electrons are emitted at a lower temperature than the primary electrons near the boundary, the emitted flux has a lower energy than the bulk flux. Hence, the average electron temperature  drops at the boundary, resulting in a spike in the mean energy when the emitted electrons meet the flow from the bulk. 
\par
\begin{figure}[h]   
  \centering
  \begin{subfigure}[t]{0.45\textwidth}
    \centering
    \includegraphics[scale=0.24]{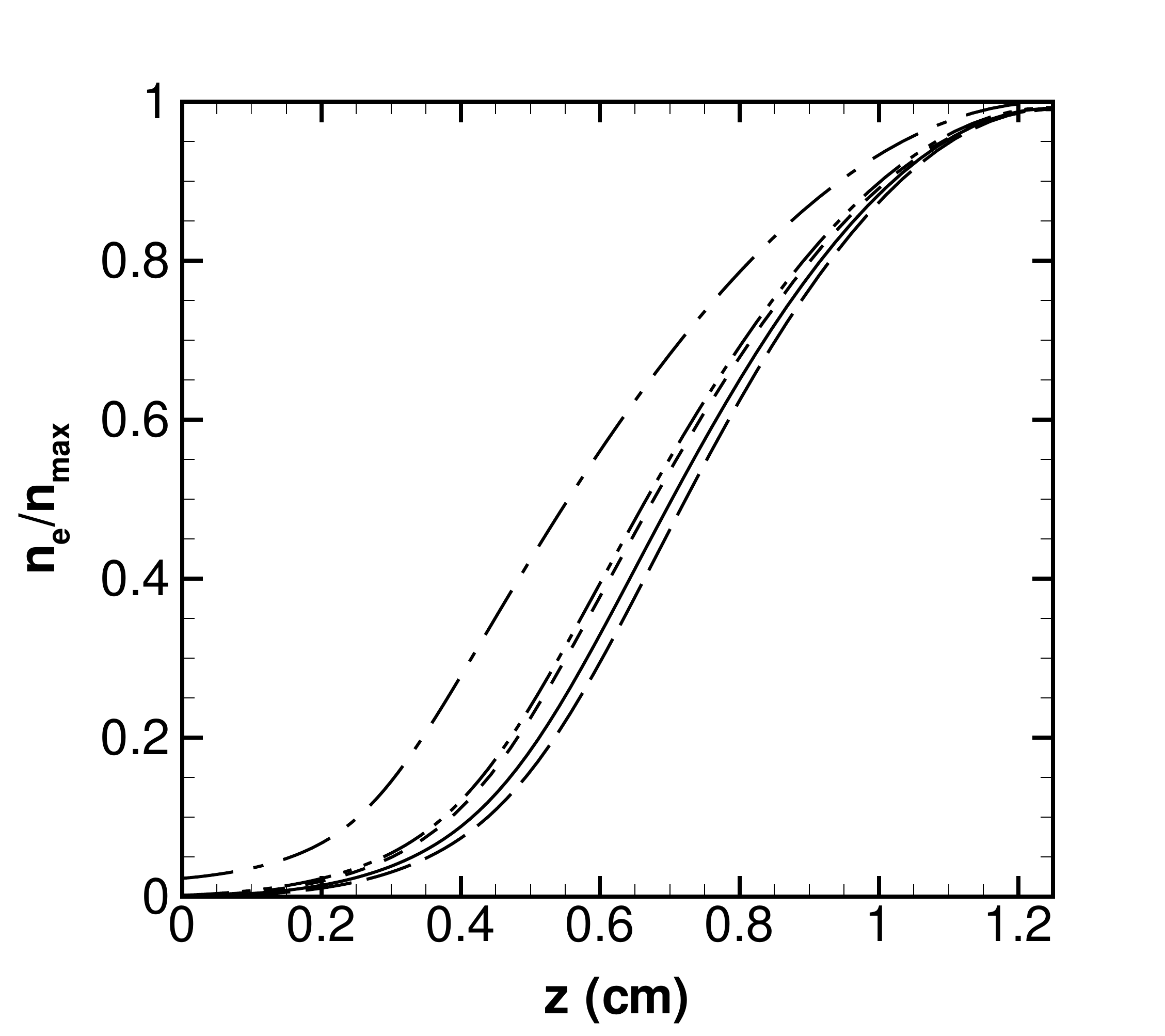}
    \caption{}
    \label{fig:ne_normal}
  \end{subfigure}
  ~
  \begin{subfigure}[t]{0.45\textwidth}   
    \centering
    \includegraphics[scale=0.24]{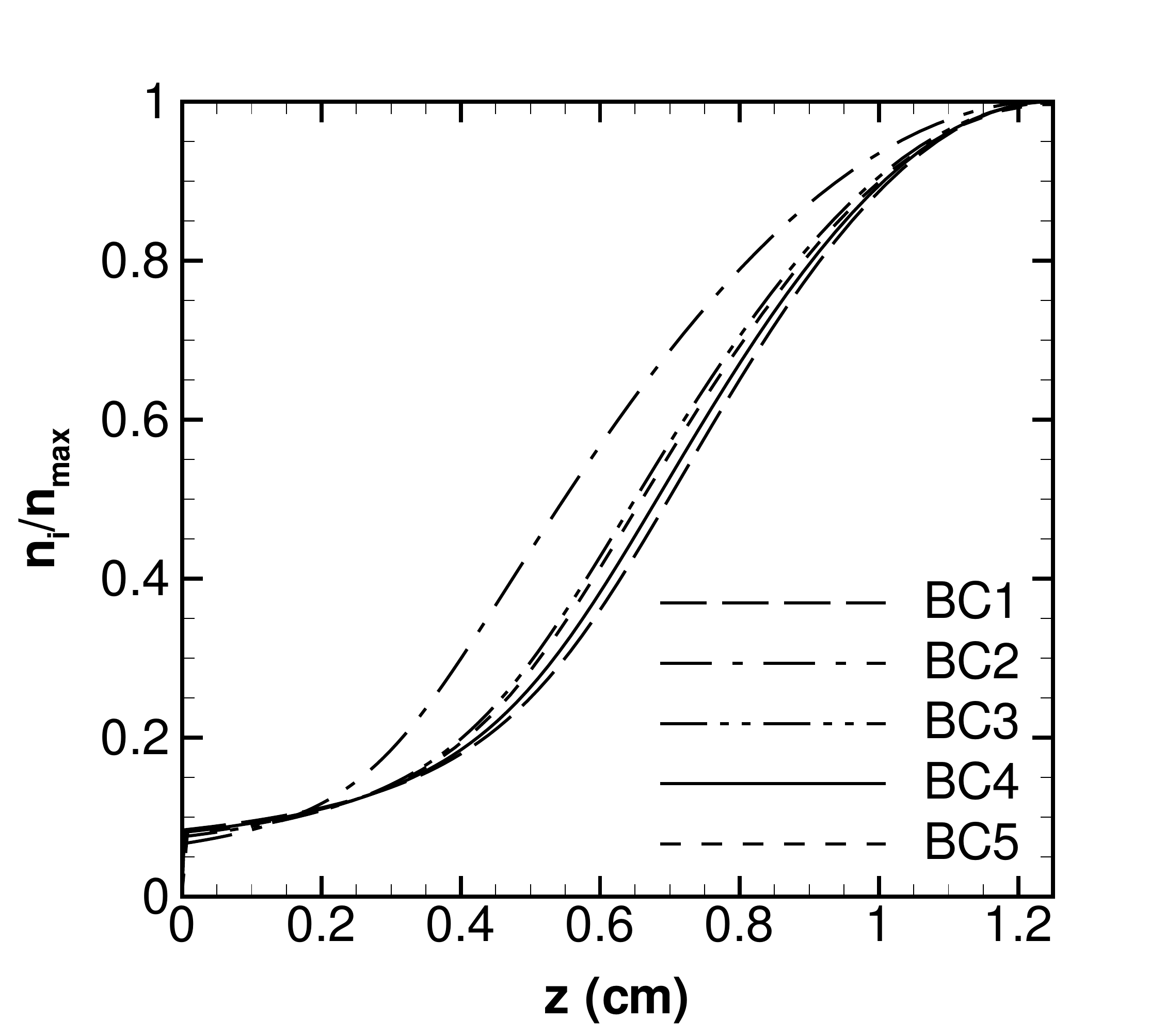}
    \caption{}
    \label{fig:ni_normal}
  \end{subfigure}
  \caption{Spatial variation of (a) normalized electron number density and (b) normalized ion number density.}
  \label{fig:1D_Normalized}
\end{figure}

 The maximum bulk plasma density, which is at the middle of the domain, is tabulated for  various BCs in Table \ref{tab:BC_ne1D}. 
To examine the effect of the BC on the sheath, the electron and ion number densities were normalized by the maximum plasma density with the results  shown in Fig.~\ref{fig:1D_Normalized}. BC1 produces the largest sheath as a result of neglecting the flux to the boundary. While the sheath widths for BC1 and BC4 only vary by 3\%,  the difference is  10\%  in the bulk plasma density.  BC2 produces the smallest sheath width and is the case where the normalized electron number density at the boundary is more pronounced than the other cases. This behavior is associated with a bulk density almost 200\% higher than that in other cases. 
Neglecting the diffusion flux caused BC3 to have a smaller sheath than BC4 and a 26\% higher bulk density. 
Inclusion of SEE does not suggest a significant impact on the sheath width. 

%
\subsection{Two-dimensional axisymmetric cylindrical configuration results}
\par A cylindrical plasma reactor was modeled to investigate the influence of the different number and energy density BCs in an axisymmetric two-dimensional plasma. The plasma in this configuration is enclosed  between two electrodes at the top and bottom of the domain and a dielectric lateral wall. Two groups of cases were modeled.  In the first group, which was for the investigation of the electrode BC impact, the BC on the dielectric wall was identical (BC4, see Tab \ref{tab:BClabels})  among the cases while the number and energy density BC applied to the electrodes was different (BC1\textendash5, see Tab \ref{tab:BClabels}). In the second group, which was for the investigation of the lateral BC impact, the electrode BC was identical (BC4, see Tab \ref{tab:BClabels}) among the cases while the number and energy density BC applied to the dielectric wall was different (BC1\textendash5, see Tab \ref{tab:BClabels}).
\subsubsection{Impact of the electrode boundary condition}
\begin{figure}[h]   
  \centering
  \begin{subfigure}[t]{0.45\textwidth}
    \includegraphics[scale=0.40]{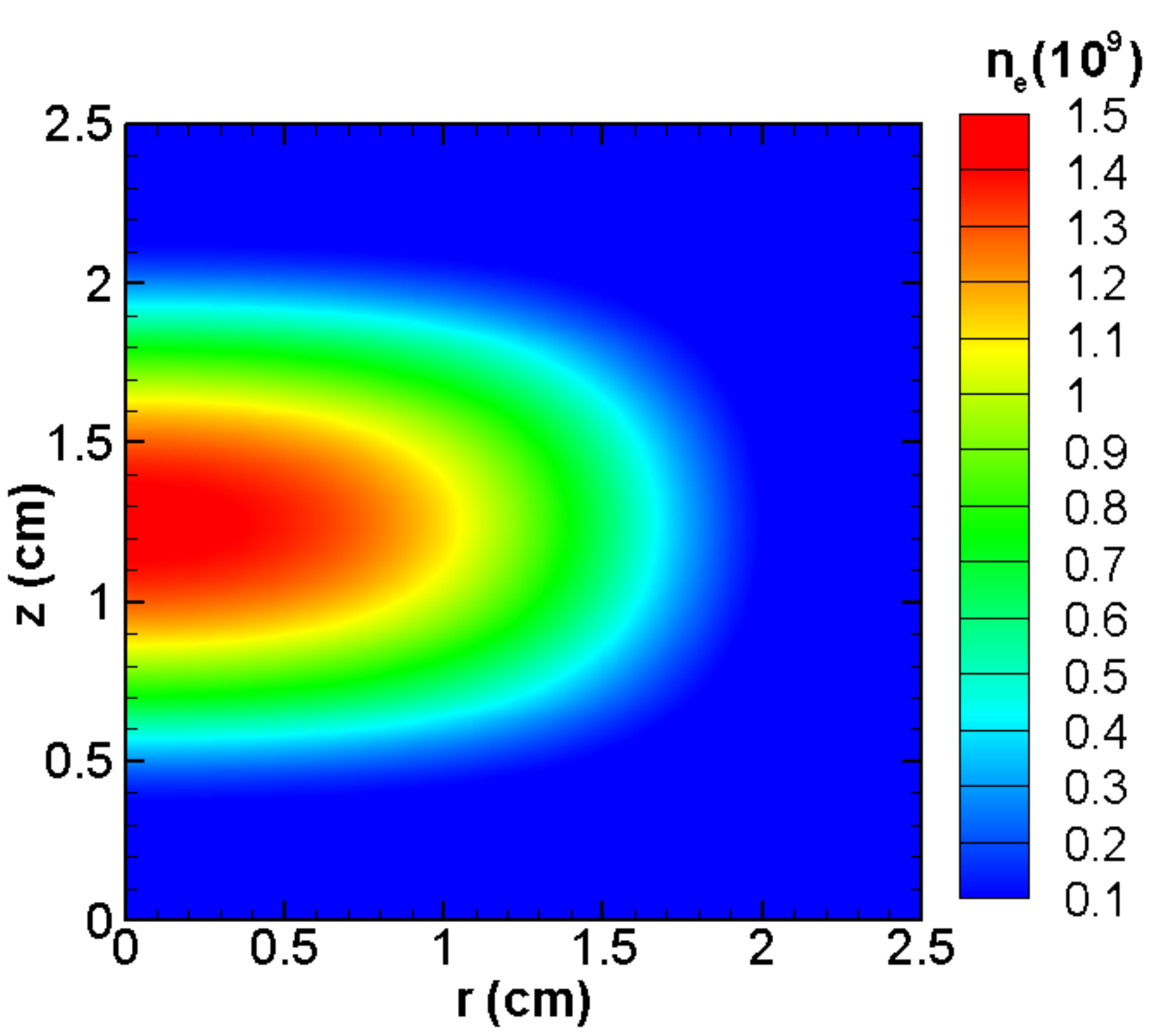}
    \caption{}\textbf{}
    \label{fig:2Dn02}
  \end{subfigure}
  ~
  \begin{subfigure}[t]{0.45\textwidth}   
    \centering
    \includegraphics[scale=0.40]{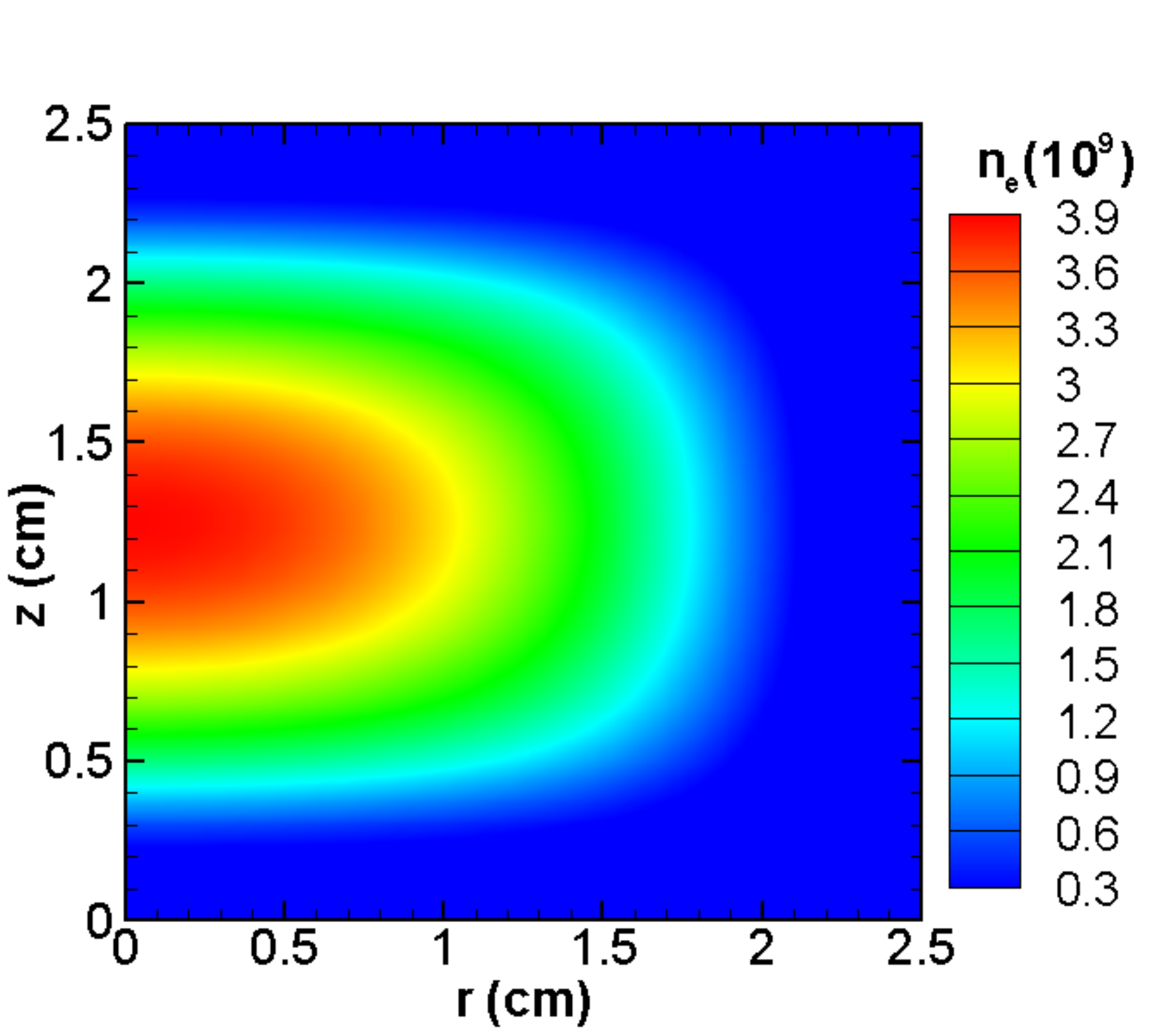}
    \caption{}
    \label{fig:2Ddrift2}
  \end{subfigure}
  \caption{Contourplots of RF period-averaged electron number density (in 10$^9$ cm$^{-3}$) with  (a) BC1; and (b) BC2 as the electrode boundary condition. 
  The lateral dielectric wall located at $r=2.5$ is set to BC4 in both subfigures.}
  \label{fig:2D_ne_2}
\end{figure}

Here, the axisymmetric cylindrical plasma reactor was modeled using the different sets of BCs, shown in Table \ref{tab:BClabels}, for the electrodes but the dielectric wall was set to BC4 in all cases. 
All BC kinds produced qualitatively similar results, with BC1 producing the lowest plasma density and BC2 producing the highest in the bulk of plasma. The RF-averaged electron number density contours are shown in  Fig.~\ref{fig:2D_ne_2} for BC1 and BC2. 

\par
\begin{figure}[h]   
  \centering
  \begin{subfigure}[t]{0.45\textwidth}
    \includegraphics[scale=0.40]{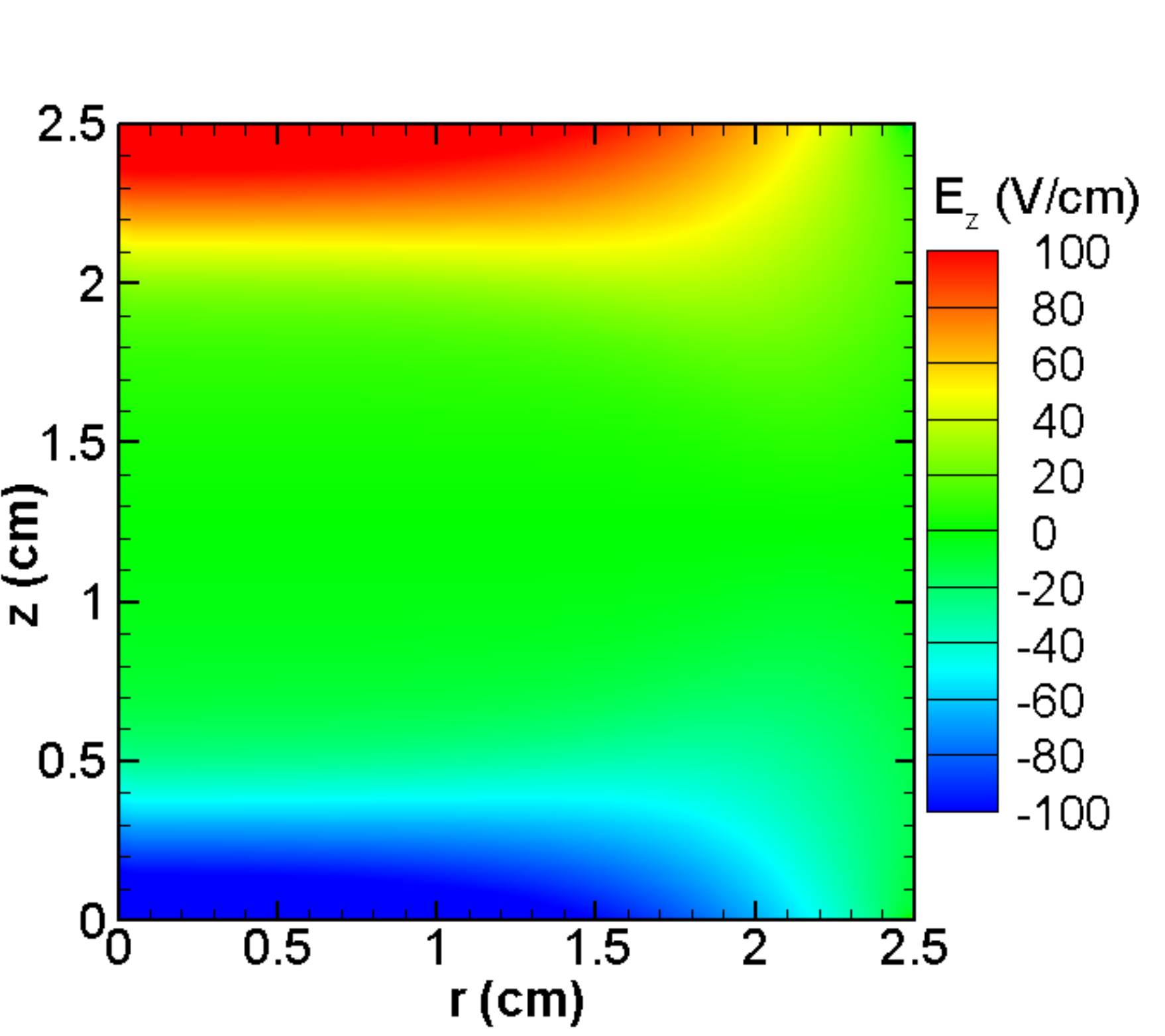}
    \caption{}
    \label{fig:Ez_plot}
  \end{subfigure}
  ~
  \begin{subfigure}[t]{0.45\textwidth}   
    \centering
    \includegraphics[scale=0.40]{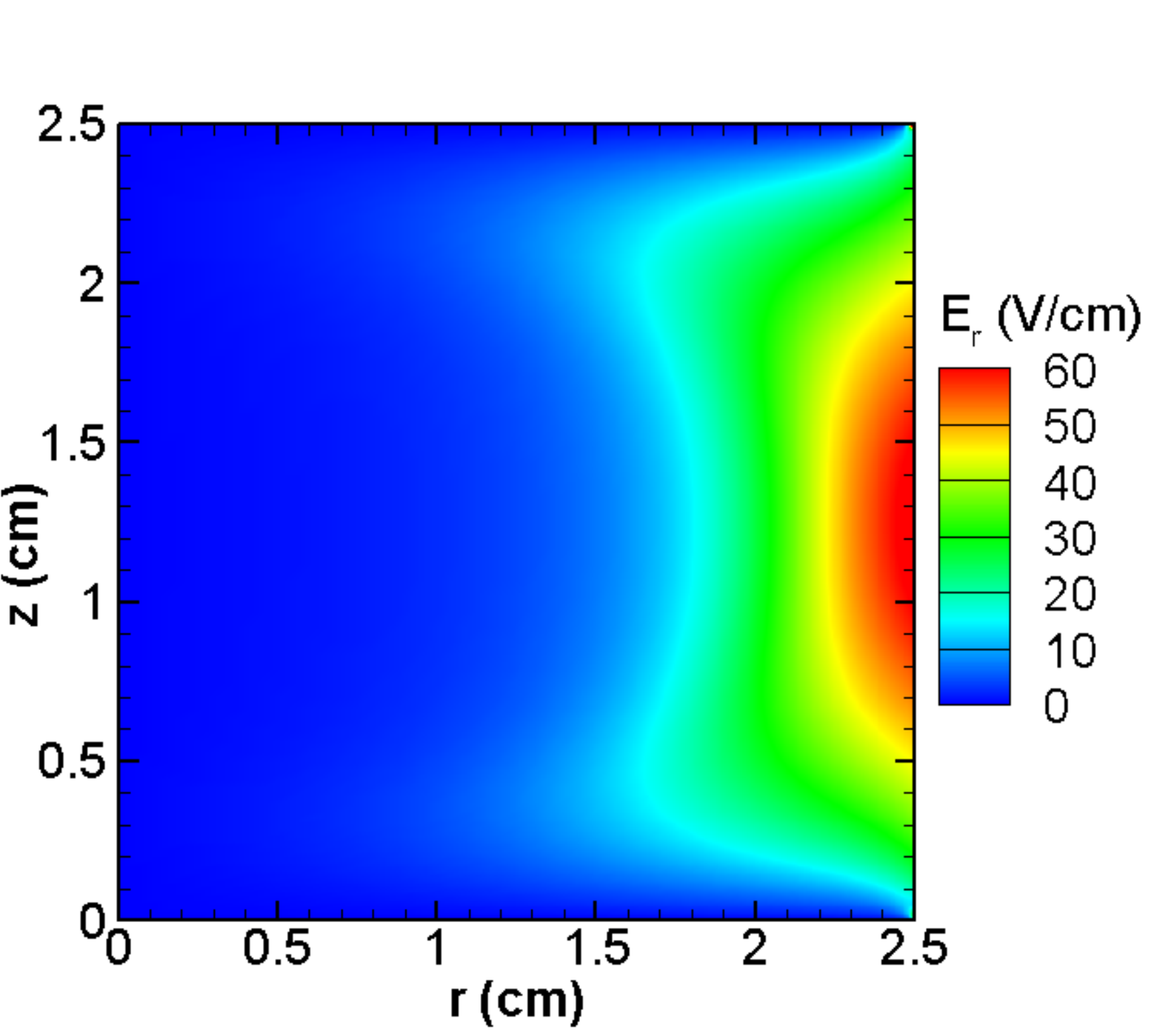}
    \caption{}
    \label{fig:Er_plot}
  \end{subfigure}
  \caption{Contour plots of RF period-averaged electric field in V/cm, BC4 is used for both the electrodes and the lateral wall; (a) vertical electric field and (b) radial electric field. 
  }
  \label{fig:2D_efield}
\end{figure}

 To determine whether the ion extrapolation BC can be used, it is necessary to check the electric field along the boundary since ion extrapolation is only valid when the ions are flowing towards the boundary. 
An examination of the electric field, shown in Fig.~\ref{fig:2D_efield} reveals that it is positive on the upper electrode and the wall, and negative on the lower electrode. 
However, the electric field magnitude near the corners is low. An examination of the time variation showed that there are times in the RF cycle where the electric field in the corners of the domain changes direction. Therefore, the ion extrapolation boundary condition cannot be applied for every point on the boundaries and it is necessary to check the local value of the electric field before using ion extrapolation.
Applying an ion extrapolation BC did not have a significant effect on the plasma, compared to the base case. 

\par
\begin{figure}[h]   
  \centering
  \begin{subfigure}[t]{0.45\textwidth}
    \centering
    \includegraphics[scale=0.25]{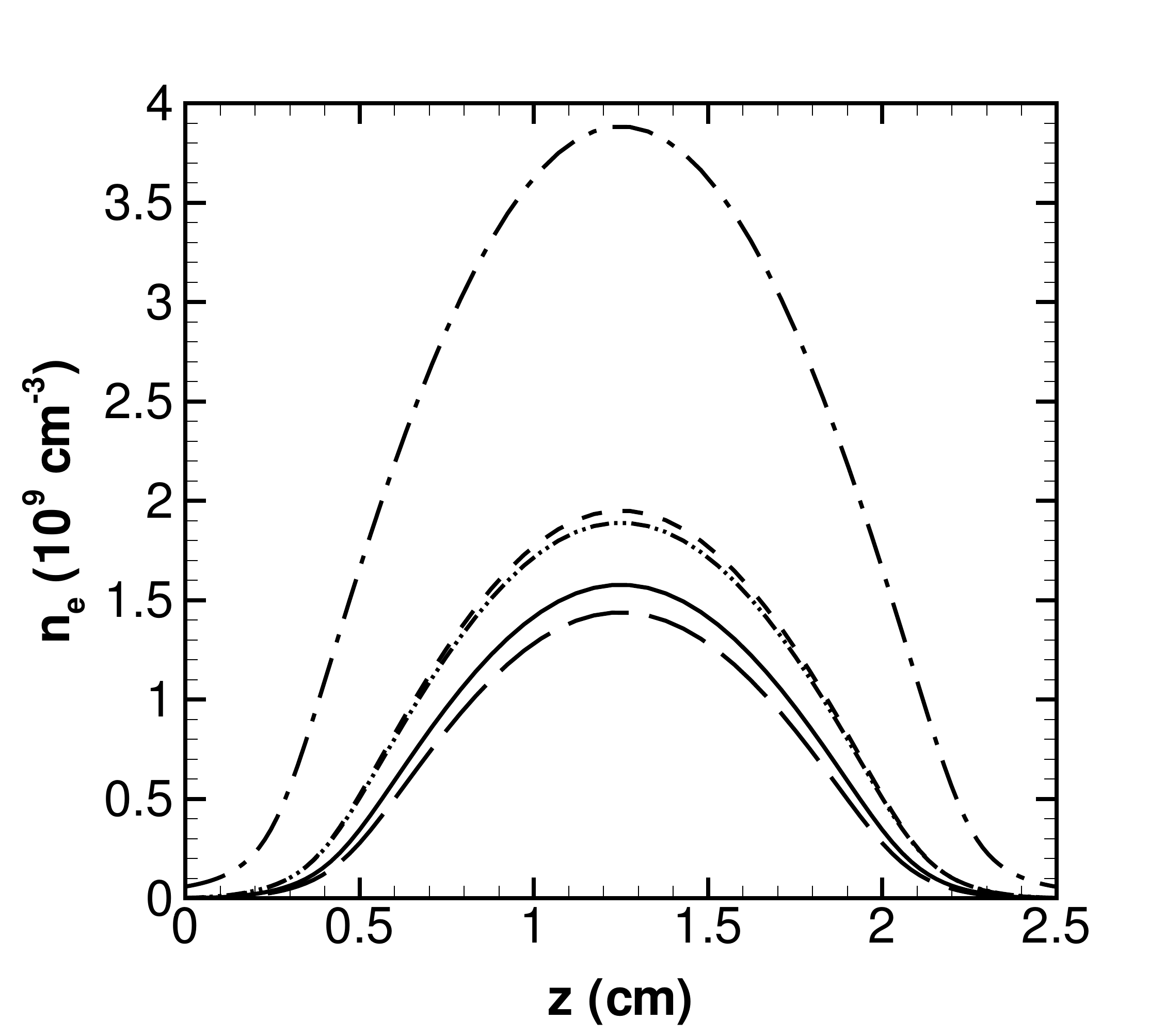}
    \caption{}
    \label{fig:ne_axis}
  \end{subfigure}
  ~
  \begin{subfigure}[t]{0.45\textwidth} 
    \centering
    \includegraphics[scale=0.25]{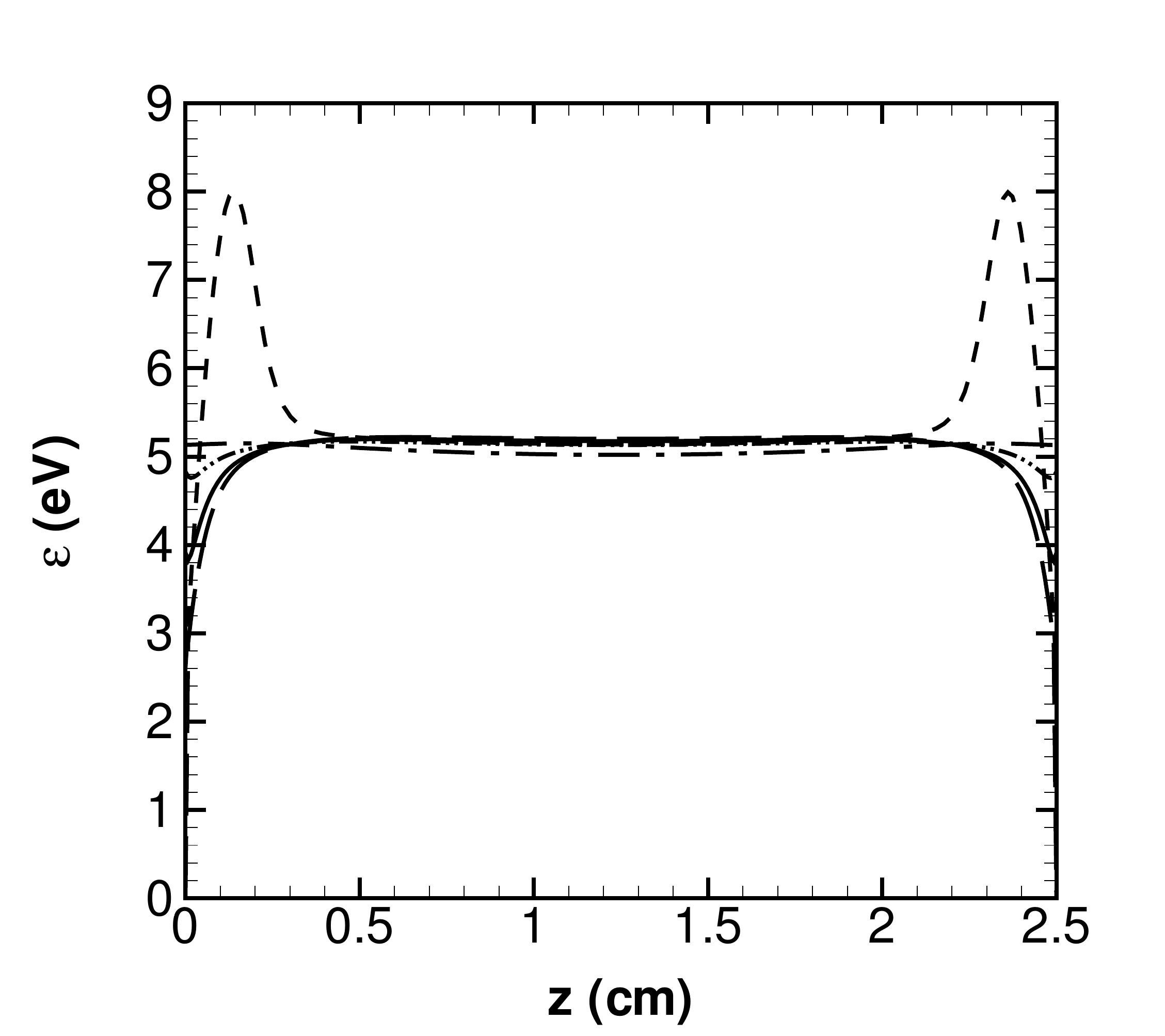}
    \caption{}
    \label{fig:eps_axis}
  \end{subfigure}
  
  \begin{subfigure}[h]{0.45\textwidth}
    \centering
    \includegraphics[scale=0.25]{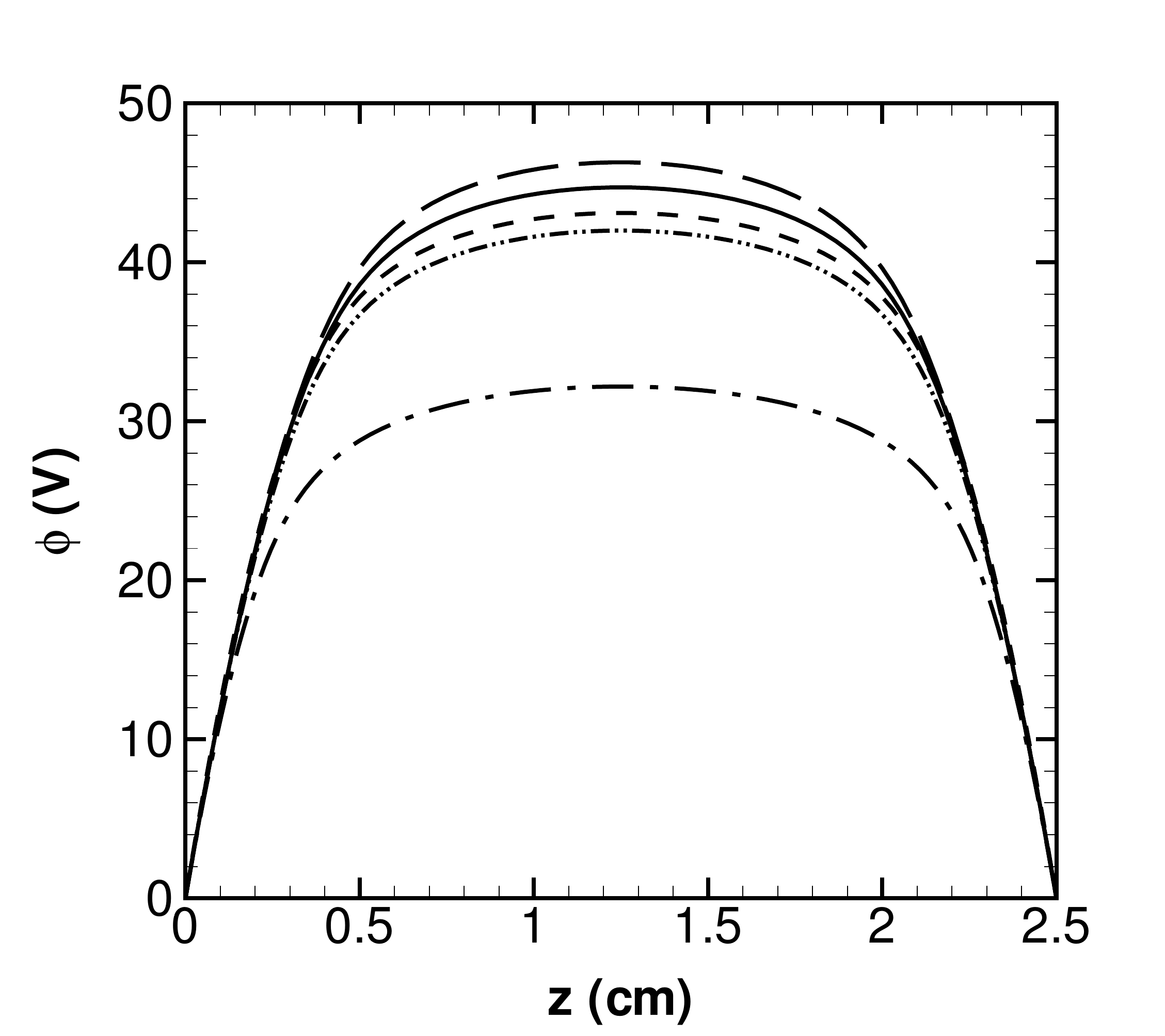}
    \caption{}
    \label{fig:potential_axis}
  \end{subfigure}
  ~
  \begin{subfigure}[h]{0.45\textwidth}  
    \centering
    \includegraphics[scale=0.25]{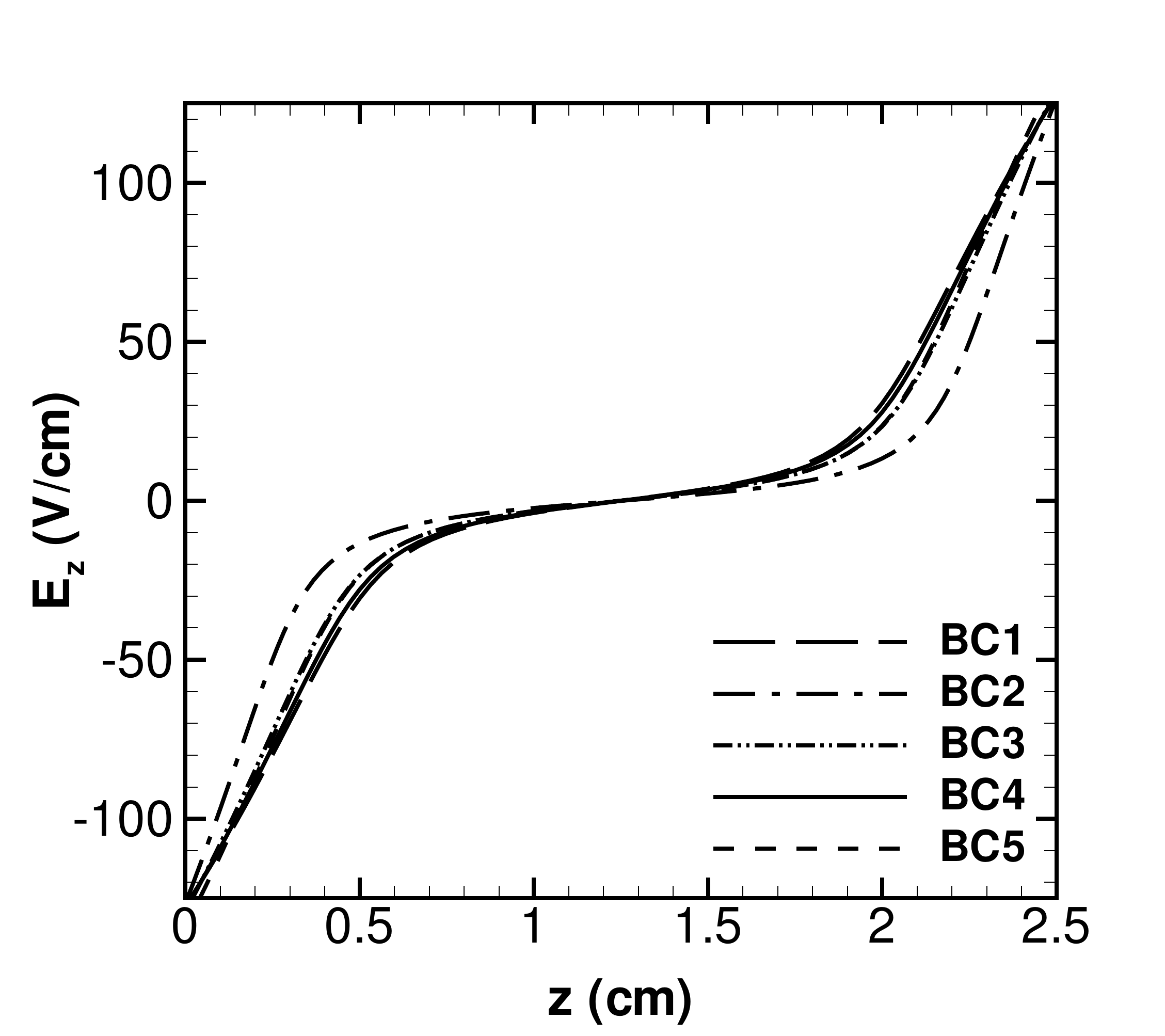}
    \caption{}
    \label{fig:Ez_axis}
  \end{subfigure}
  \caption{Spatial variation of RF-averaged plasma variables in the vertical direction at $r=0$ cm for the different electrode boundary conditions: (a) electron number density (in $10^9$ cm$^{-3}$); (b) mean electron energy (in eV); (c) electric potential (in V); and (d) vertical electric field (in V/cm).} 
  \label{fig:2DresultsZ_2}
\end{figure}

 The spatial variation of the RF-averaged plasma variable along the axisymmetric line ($r=0$) is plotted in Fig.~\ref{fig:2DresultsZ_2}. As seen in Fig.~\ref{fig:ne_axis}, BC2 has a substantially higher plasma density than the other cases, which is due to a lower rate of electron loss at the surface in this case. BC2 also produces a lower electric potential than the other cases, as seen in Fig.~\ref{fig:potential_axis}. 
The SEE, which is included in BC5, caused a 24\% increase in the plasma density (Fig.~\ref{fig:ne_axis}) and also caused a spike in the mean energy near the electrodes (Fig.~\ref{fig:eps_axis}), compared to the base case. 

\par
\begin{figure}[h]   
  \centering
  \begin{subfigure}[t]{0.45\textwidth}
    \centering
    \includegraphics[scale=0.25]{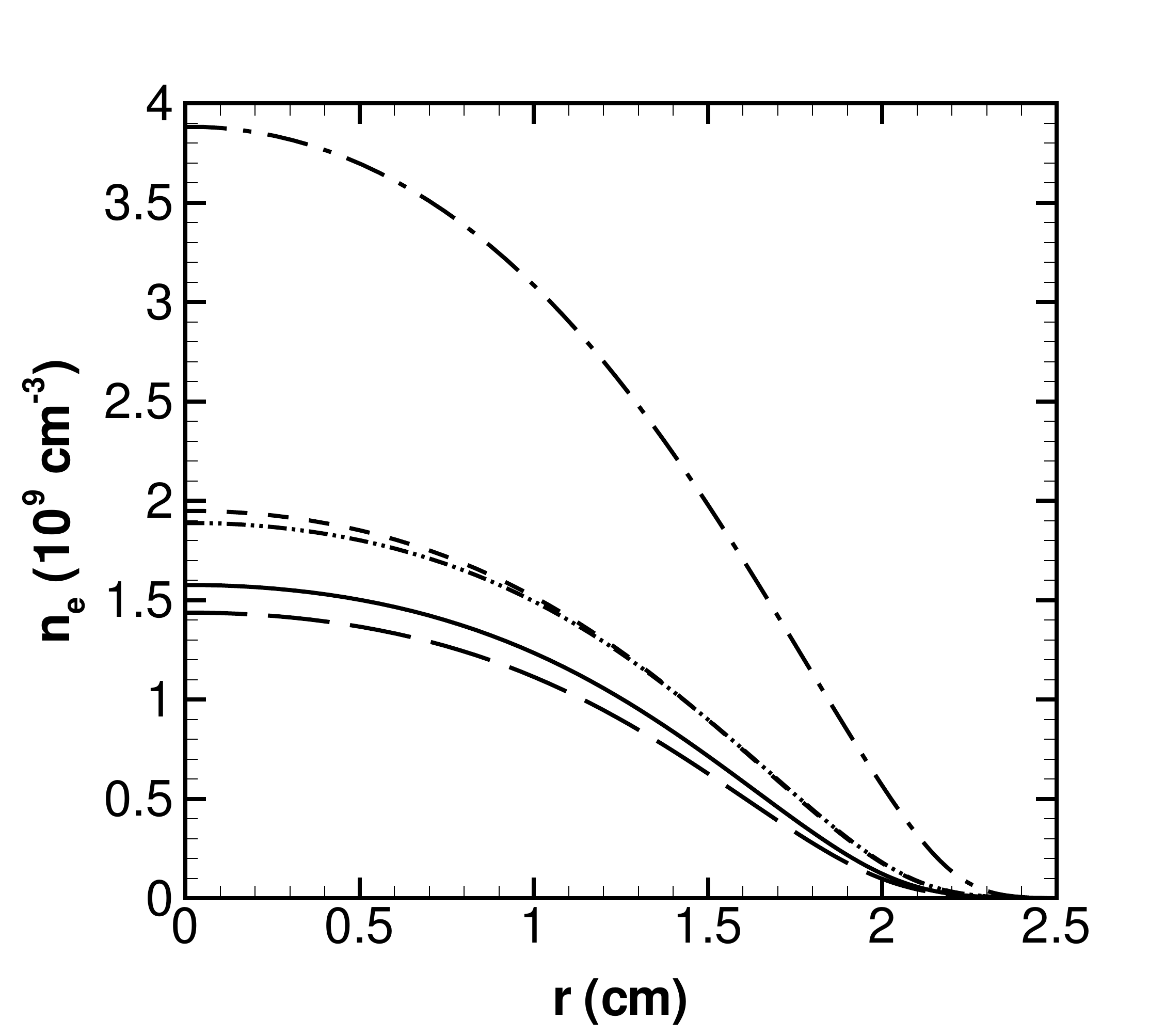}
    \caption{}
    \label{fig:ne_radial2}
  \end{subfigure}
  ~
  \begin{subfigure}[t]{0.45\textwidth} 
    \centering
    \includegraphics[scale=0.25]{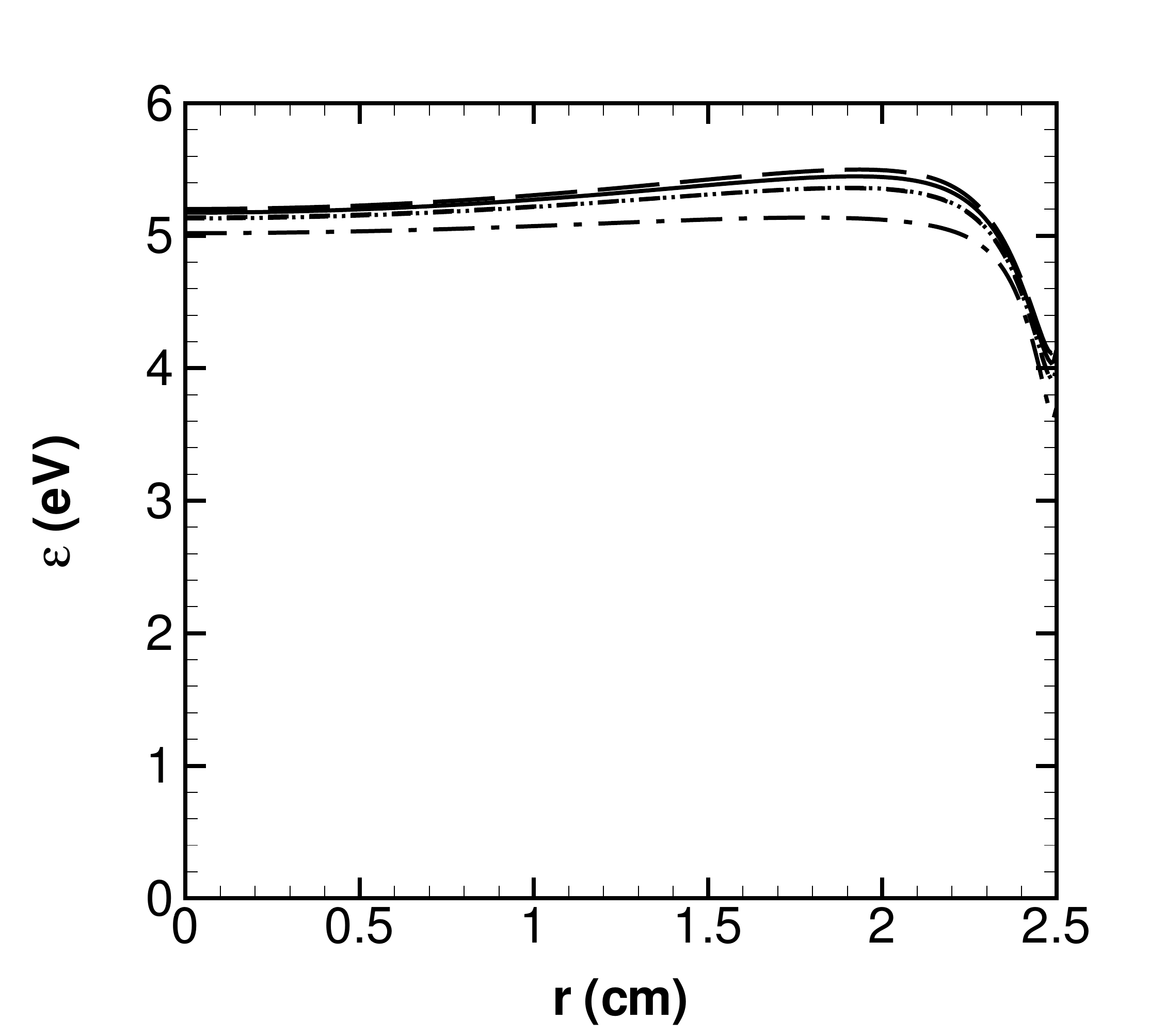}
    \caption{}
    \label{fig:eps_radial2}
  \end{subfigure}
  
  \begin{subfigure}[h]{0.45\textwidth}
    \centering
    \includegraphics[scale=0.25]{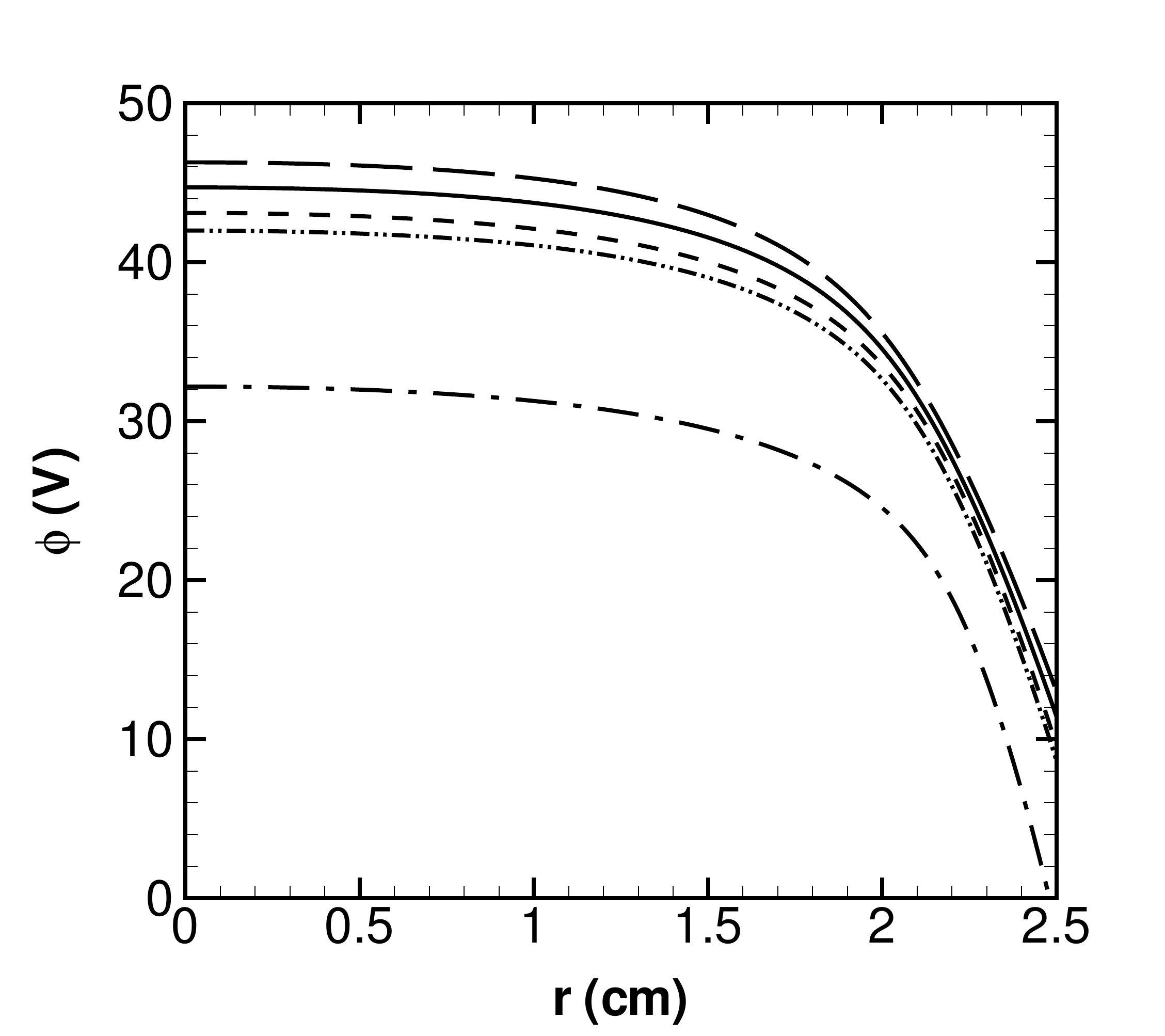}
    \caption{}
    \label{fig:potential_radial2}
  \end{subfigure}
  ~
  \begin{subfigure}[h]{0.45\textwidth}  
    \centering
    \includegraphics[scale=0.25]{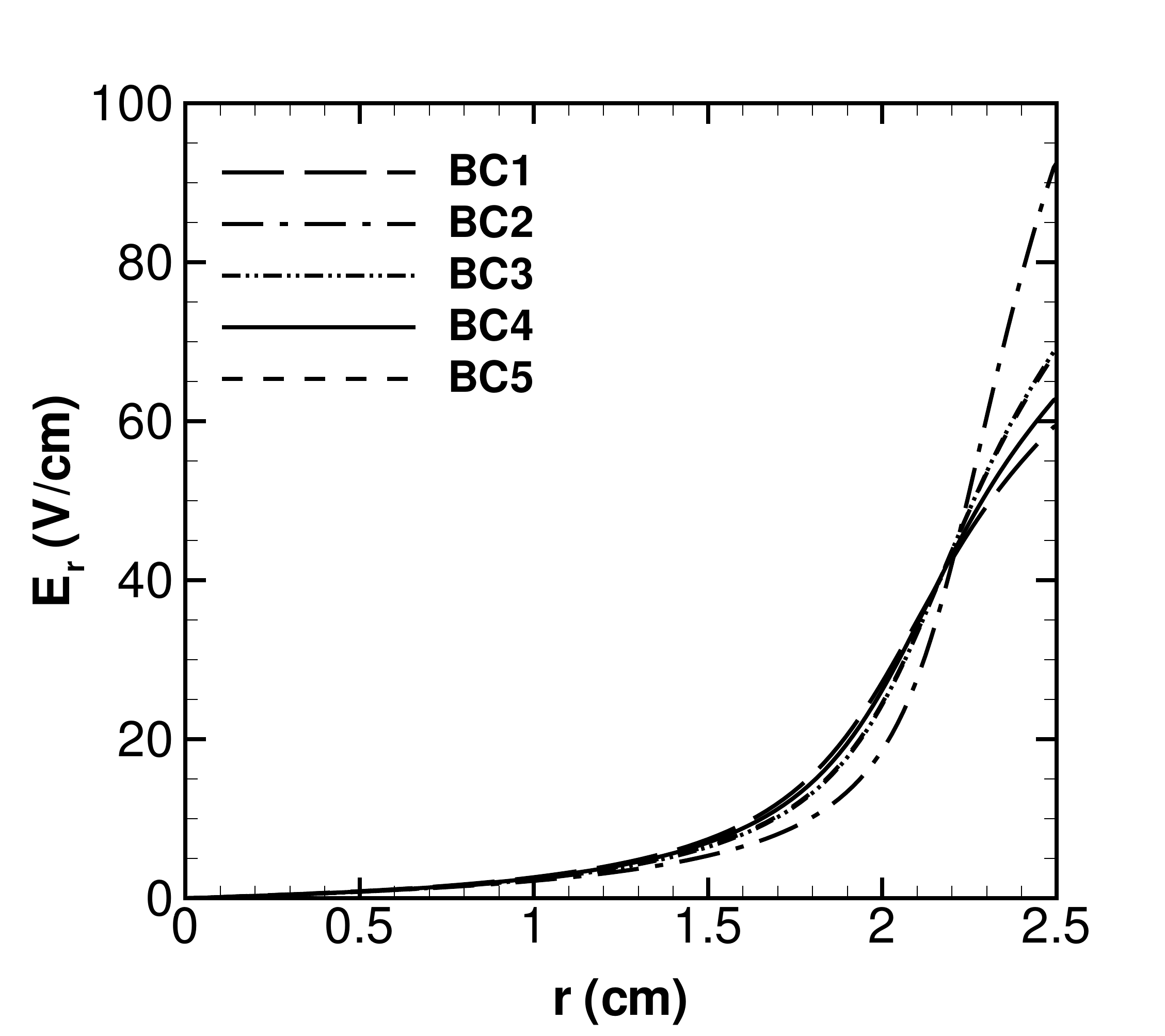}
    \caption{}
    \label{fig:Er_radial2}
  \end{subfigure}
  \caption{Spatial variation of RF-averaged plasma variables in the radial direction at $z=1.25$ cm, for the different electrode boundary conditions: (a) electron number density (in $10^9$ cm$^{-3}$); (b) mean electron energy (in eV); (c) electric potential (in V); and (d) radial electric field (in V/cm).} 
  \label{fig:2Dresults_2}
\end{figure}

 The radial variation of the plasma variables at $z=1.25$cm is plotted in Fig.~\ref{fig:2Dresults_2}. 
Since the dielectric wall BC was kept constant and did not include secondary emission, BC5 did not produce a spike in the mean energy near the wall, only near the electrodes. Also, since the electrode BC affected the bulk plasma density, it also affected the flux to the dielectric wall. Therefore, the electric field at the wall was affected by the choice of electrode BC. The magnitude of the electric field at the wall, as seen in Fig.~\ref{fig:Er_radial2}, was lowest for BC1, which had the lowest plasma density, and highest for BC2, which had the highest plasma density.

\par The bulk plasma density calculated at the center of the reactor $z=1.25$ cm and $r=0$ cm, is given in Table \ref{tab:BC_neElectrode} for the various cases. At the center, the RF-averaged densities are maximum, compared to the rest of the domain.  
It is seen in this table that the zero number density BC1 reduced the maximum plasma density by 9\% compared to the baseline case, due to the larger sheath that occurs in BC1. 
BC3 had a 20\% higher maximum plasma density than BC4 because diffusion flux to the surface is lacking in BC3. Secondary emission was of the same order as the diffusion flux with 
a 24\% higher maximum plasma density for BC5 compared to BC4. 
\subsubsection{Impact of the dielectric wall boundary condition} 
Here, BC4 was used for electrodes and different BCs (Tab.~\ref{tab:BClabels}) were used for the dielectric wall. 

\begin{figure}[h]   
  \centering
  \begin{subfigure}[t]{0.45\textwidth}
    \includegraphics[scale=0.40]{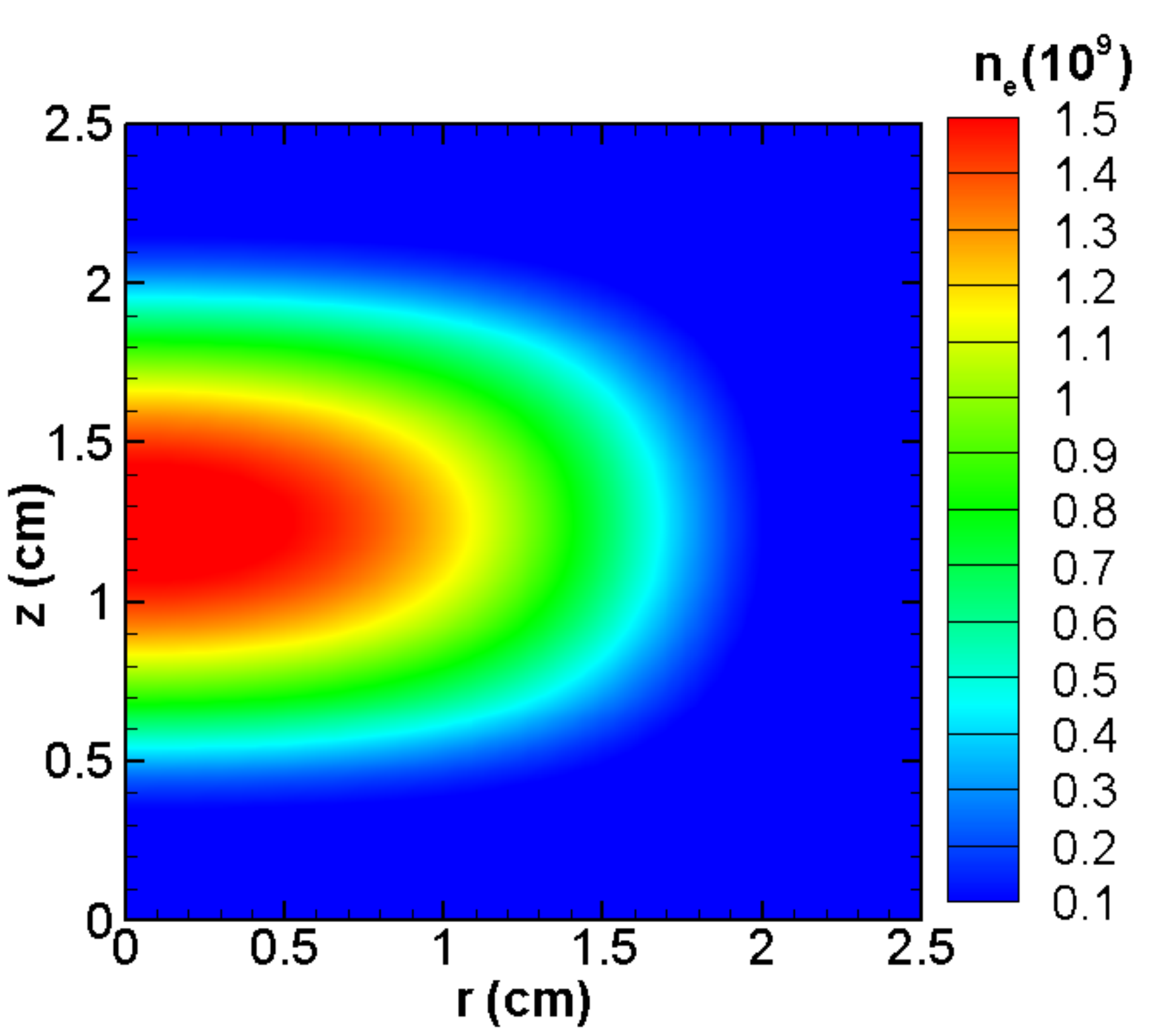}
    \caption{}
    \label{fig:2Dn0}
  \end{subfigure}
  ~
  \begin{subfigure}[t]{0.45\textwidth}   
    \centering
    \includegraphics[scale=0.40]{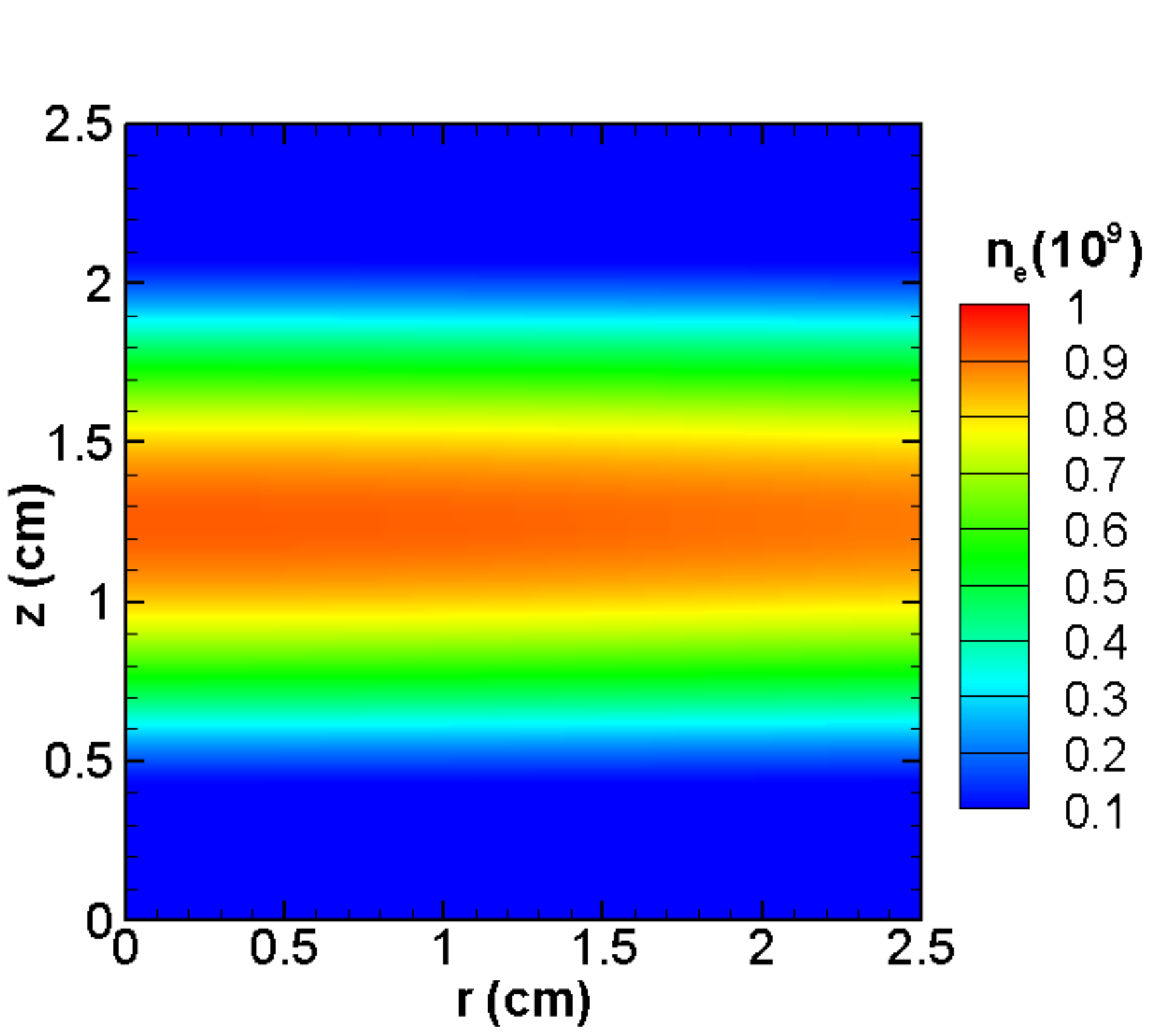}
    \caption{}
    \label{fig:2Ddrift}
  \end{subfigure}
  \caption{Contour plots of RF period-averaged electron number density (in 10$^9$ cm$^{-3}$) for a dielectric wall (right boundary) using (a) BC1 and (b) BC2. The results produced by BC3, BC4, and BC5 were similar to BC1.}
  \label{fig:2D_ne}
\end{figure}

All cases, except the one using BC2 on the dielectric wall, produced similar results with the bulk plasma density varying less then 2\% among the cases (Tab.~\ref{tab:BC_neWall}). As seen in Fig.~\ref{fig:2D_ne}, BC2, which is the pure drift BC, did not produce a sheath at the wall, although there is a slight reduction in the density at the wall. 
The lack of sheath in the BC2 case for the dielectric wall, is a result of setting the electron flux to zero when the electric field is either zero or directed out of the domain. When this BC was applied to the electrodes, it caused an increase in the electron number density in the sheath and a reduction in the sheath width. This behavior is correlated with the strength of the electric field. A weaker field results in an even smaller sheath. For a dielectric wall, the electric field depends on the number density through the development of a wall charge, determined by eq.~(\ref{eq:dielectric}). When BC2 is used for ${\bm\Gamma}_e\cdot{\bm n}$, it prevents the development of a wall charge. This results in zero electric field, zero electron flux and zero wall charge. Setting the electric field and the fluxes to zero prevents the development of a sheath. Therefore, there is no sheath along the majority of the wall for BC2. The interaction between the electrodes and the wall near the corners of the domain causes the radial electric field in the corners of the domain to be non-zero which prevents a pure one-dimensional plasma. 

\par
\begin{figure}[h]   
  \centering
  \begin{subfigure}[t]{0.45\textwidth}
    \centering
    \includegraphics[scale=0.25]{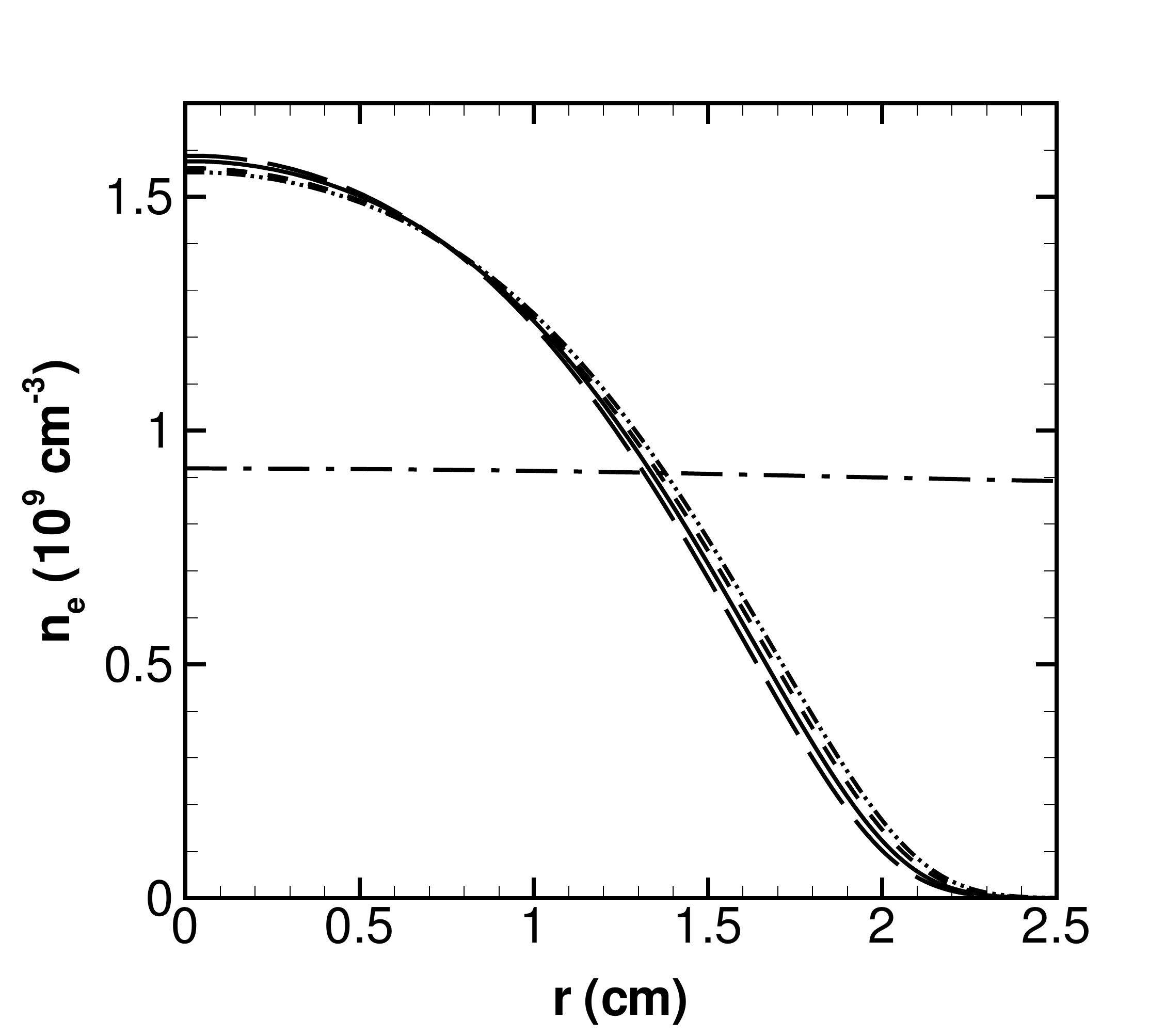}
    \caption{}
    \label{fig:ne_radial}
  \end{subfigure}
  ~
  \begin{subfigure}[t]{0.45\textwidth} 
    \centering
    \includegraphics[scale=0.25]{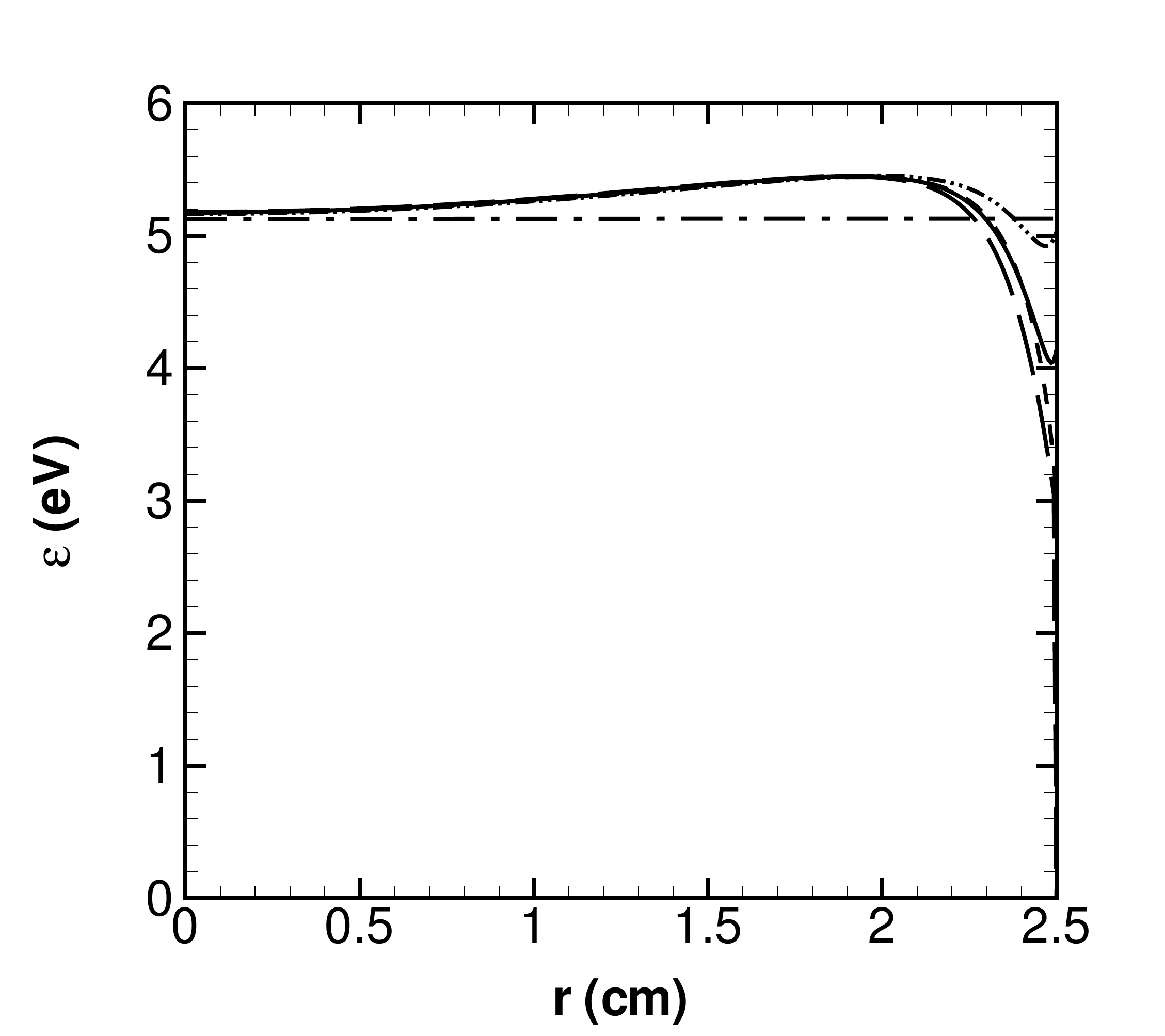}
    \caption{}
    \label{fig:eps_radial}
  \end{subfigure}
  
  \begin{subfigure}[h]{0.45\textwidth}
    \centering
    \includegraphics[scale=0.25]{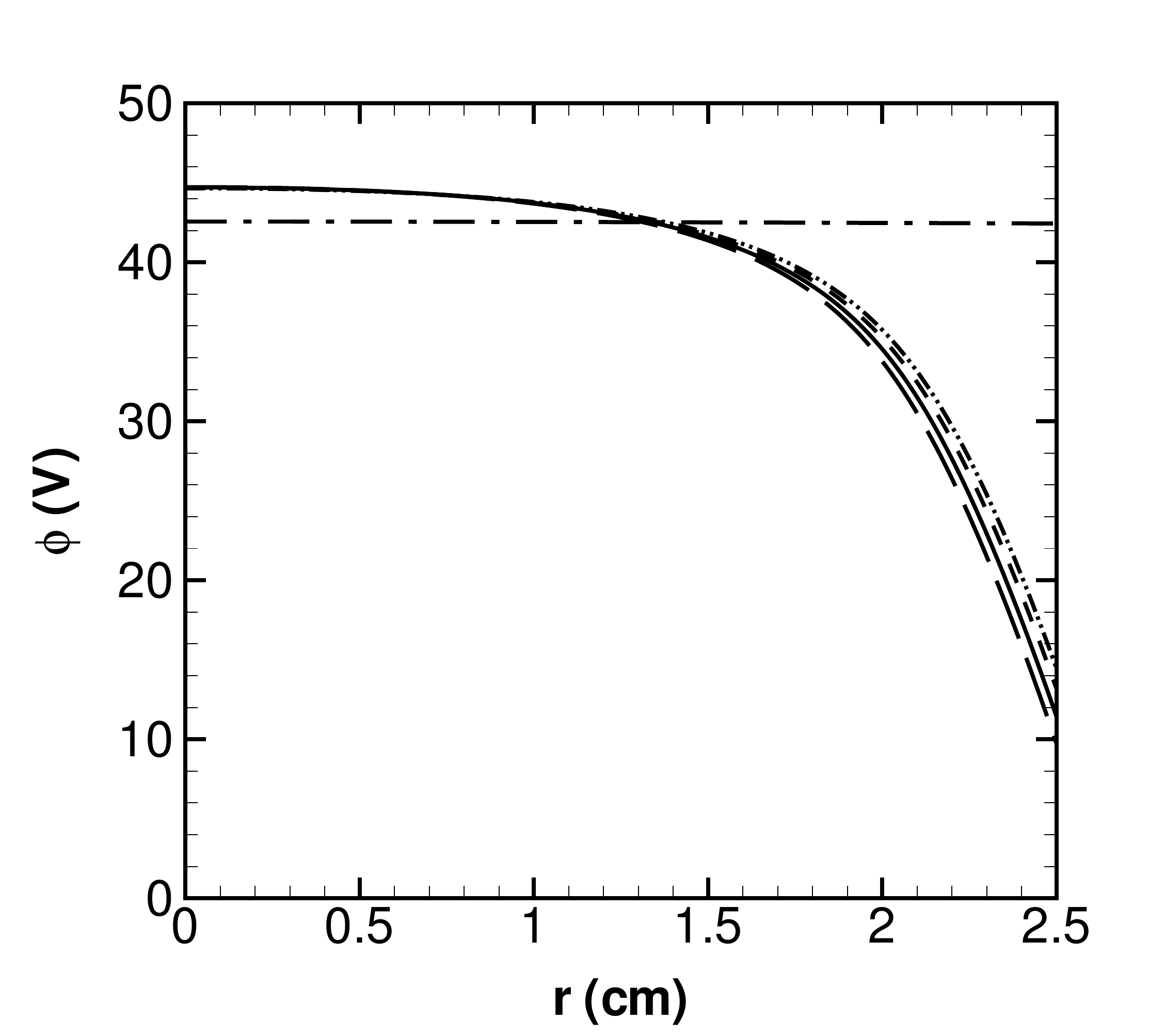}
    \caption{}
    \label{fig:potential_radial}
  \end{subfigure}
  ~
  \begin{subfigure}[h]{0.45\textwidth}  
    \centering
    \includegraphics[scale=0.25]{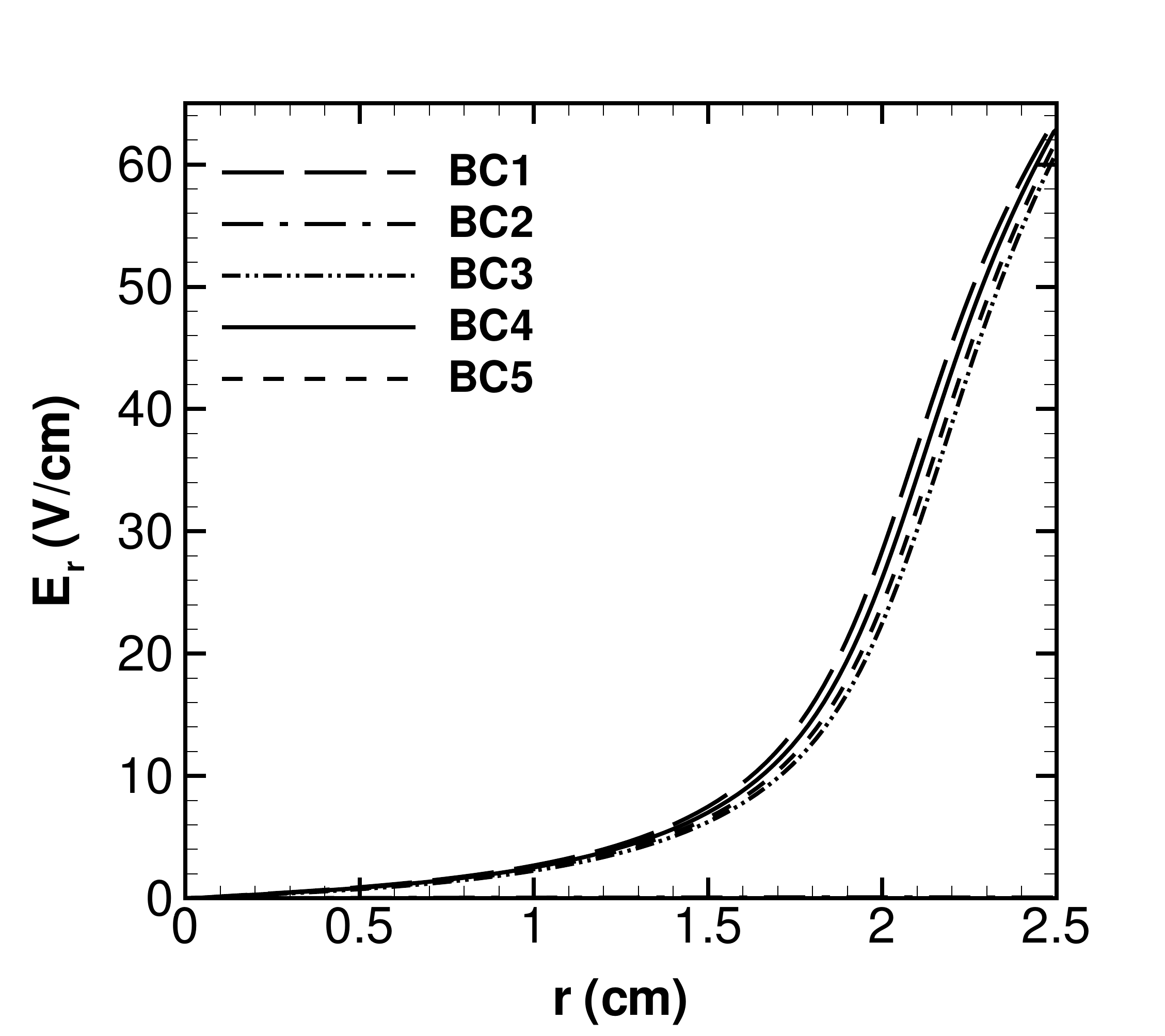}
    \caption{}
    \label{fig:Er_radial}
  \end{subfigure}
  \caption{Spatial variation of RF-averaged plasma variables in the radial direction at z=1.25 cm, for the different wall boundary conditions: (a) electron number density (in $10^9$ cm$^{-3}$); (b) mean electron energy (in eV); (c) electric potential (in V); and (d) radial electric field (in V/cm).} 
  \label{fig:2Dresults}
\end{figure}

 The plasma variables are plotted against $r$ at $z=1.25$ cm in Fig.~\ref{fig:2Dresults}. 
The lack of a sheath on the dielectric wall when using BC2, is evident in all  variables. BC3, which includes the thermal flux but neglects the diffusive flux, produced a higher electron mean energy at the wall; however, that is the only significant difference between BC3-5 and BC1. The effect of the SEE from the dielectric wall is not significant in this setup. BC5 shows a 1\% decrease in the bulk plasma density, compared to BC4. 
The electric field at the dielectric wall is a function of the ion and electron flux to the wall, which suggests that the change of the number density BC should have a significant effect on the electric field. However,  Fig.~\ref{fig:Er_radial} shows that all the BCs except BC2 produce a RF-averaged radial electric field of 60~64 V/cm at the boundary. The use of BC2 results in zero radial electric field throughout the domain as evident in \ref{fig:2Ddrift}. 
This indicates that while the different BCs results in different values for the electron and ion fluxes at the wall, the increase or decrease in the electron flux is balanced out by a similar increase or decrease in the ion flux with the exception of BC2. Hence, the total current to the wall is similar for all  BCs except BC2. 

\section{Conclusions}  
\label{sec:Conclusions}
The change of the BC kinds implemented for the electrodes had a significant impact on the plasma. For the ions, the mobility was larger than the diffusion which caused the ions to be dominated by the drift motion. Therefore, since the electric field in an RF plasma generally causes the ions to flow out of the domain, the ion BC should not have a significant effect on the plasma. This was verified by the fact that using an ion extrapolation BC did not have a significant effect on the plasma. 
        	
The electrons have a larger diffusion and are not dominated by the drift motion in the same fashion as the ions. This behavior means the electron BC should have a significant effect and it is important to account for all the contributions to the electron flux at the boundary. Neglecting the flux and using the zero number density BC produces a 10\% lower bulk plasma density than when the BC is based on a complete description of the flux to the electrode. Neglecting the thermal and diffusion fluxes and including only the drift towards the surface increases the bulk plasma density by a factor between two and three, compared to all the other BC’s. Including both the drift and the thermal fluxes but neglecting the diffusion flux resulted in a 20-30\% increase in the bulk plasma density. Including secondary electron emission, with a coefficient of $\gamma=0.5$, resulted in a 20-25\% increase in the bulk plasma density as a result of the increased mean energy in the sheath. 

The plasma was less sensitive to the change of the dielectric wall BC than the electrode BC with the exception of the case where the electron BC at the wall was based on only the drift flux. The pure drift BC for the electrons failed to produce a sheath at the dielectric wall; hence, this BC is not recommended for the dielectric wall. The remaining BC’s for the dielectric wall produced bulk plasma densities within 2\% of each other.

\section*{Acknowledgments}
The authors acknowledge fruitful discussions with Lorin S. Matthews from Baylor University. This work was supported through the NSF/DOE Partnership in Basic Plasma Science and Engineering program with the award number PHY-1414552.

\appendix
{\section{Derivation of SEE BC}
\label{sec:seeBCDerive}}

Substituting for $\bm{\Gamma}_\alpha\cdot\bm{n}$ and $\bm{\Gamma}_\gamma\cdot\bm{n}$ from eqs. (\ref{eq:ne_bulk}) and (\ref{eq:SEEFlux}), respectively, into  $\bm{\Gamma}_{e}\cdot\bm{n}=\bm{\Gamma}_\gamma\cdot\bm{n}+\bm{\Gamma}_{\alpha}\cdot\bm{n}$ and using $n_\alpha=n_e-n_\gamma$, one obtains:
\begin{equation}
\bm{\Gamma}_{e}\cdot\bm{n}=\left[-\left(2a_e-1\right)\mu_{e}\bm{E}\cdot\bm{n}+\frac{1}{2}v_\text{th}\right]\left(n_e-n_{\gamma}\right)
-\left(1-a_e\right)\gamma\bm{\Gamma}_{i}\cdot\bm{n}.
\end{equation}

\noindent
Expanding the first term and substituting for $n_\gamma$ from eq.~(\ref{eq:ne_emit}) in the drift term gives:
\begin{equation} \begin{gathered}
\bm{\Gamma}_{e}\cdot\bm{n}=\left[-\left(2a_e-1\right)\mu_{e}\bm{E}\cdot\bm{n}+\frac{1}{2}v_\text{th}\right]n_e-\frac{1}{2}v_\text{th}n_{\gamma}
\\ +\left(2a_e-1\right)\left[\left(1-a_e\right)\frac{\gamma\bm{\Gamma}_{i}\cdot\bm{n}}{\mu_e\bm{E}\cdot\bm{n}}\right]\mu_{e}\bm{E}\cdot\bm{n}
-\left(1-a_e\right)\gamma\bm{\Gamma}_{i}\cdot\bm{n},
\end{gathered} \end{equation}

\noindent
which simplifies to the following equation after factoring out $(1-a_e)$ between the last two terms:
\begin{equation} \begin{gathered}
\bm{\Gamma}_{e}\cdot\bm{n}=\left[-\left(2a_e-1\right)\mu_{e}\bm{E}\cdot\bm{n}+\frac{1}{2}v_\text{th}\right]n_e-\frac{1}{2}v_\text{th}n_{\gamma}
\\ -2\left(1-a_e\right)^2\gamma\bm{\Gamma}_{i}\cdot\bm{n} .
\end{gathered} \end{equation}
Using the definition of $a_e$ given in eq. (\ref{eq:aei}), it can be shown that $\left(1-a_e\right)^2=\left(1-a_e\right)$, so the BC for the electrons in eq.~(\ref{eq:neBC5}) is derived. The equation for the energy density given in eq.~(\ref{eq:we_BC5}) is derived in a similar fashion.


\section*{References}

\newpage
\begin{table}
  \centering
  \caption{Plasma parameters}
  \label{tab:parameters}
  \begin{tabular}{|l|c|}
   \hline
   Parameter & Value \\
   \hline
  Electrode gap, H &  2.5 (cm) \\
  Electrode radius, R &  2.5 (cm) \\
  Applied voltage, $V_\text{RF}$ & 50 (V) \\
  Frequency, $f$ & 13.56 (MHz) \\
  Neutral gas pressure, $P_\text{gas}$ &  250 (mTorr) \\
  Ion and neutral temperature, $T_i=T_\text{gas}$ & 300 (K) \\
  Electron mobility $\mu_e P_{gas}$ & 3$\times$10$^5$ \\
  Electron diffusion $D_e P_{gas}$ & 1.2$\times$10$^6$ \\
  Ion mobility $\mu_i P_{gas}$ & 1400 \\
  Ion diffusion $D_i P_{gas}$ & 40 \\
   \hline
  \end{tabular}
\end{table}
\begin{table}
  \centering
  \caption{Number and energy density boundary conditions}
  \label{tab:BClabels}
  \begin{tabular}{|c|l|c|}
   \hline
   Label & Description & Equations\\
   \hline
  BC1 &  Zero number density (Sec. \ref{sec:ZeroN}) & \textemdash \\
  BC2 &  Pure drift towards the electrode (Sec. \ref{sec:Drift}) & (\ref{eq:DriftBC},\ref{eq:DriftBCwe}) \\
  BC3 & Thermal flux (Sec. \ref{sec:Thermal}) & (\ref{eq:BC3eq},\ref{eq:we_BC3})\\
  BC4 (baseline) & Thermal and diffusion fluxes (Sec. \ref{sec:Diff}) & (\ref{eq:ni_BC45},\ref{eq:we_BC4})\\
  BC5 & Secondary emission with $\gamma=0.5$ (Sec. \ref{sec:SecEmis}) & (\ref{eq:ni_BC45},\ref{eq:neBC5},\ref{eq:we_BC5})\\
   \hline
  \end{tabular}
\end{table}
\begin{table}
  \centering
  \caption{Effect of boundary conditions on maximum bulk plasma density in the one dimensional plasma.}
  \label{tab:BC_ne1D}
  \begin{tabular}{|c|c|c|}
   \hline
   Boundary & Bulk plasma & \% difference \\
   condition & maximum density  (cm$^{-3}$) & (vs BC4) \\
   \hline
  BC1 &  8.32$\times$10$^8$ & -10.8\%\\
  BC2 &  2.75$\times$10$^9$ & 195\%\\
  BC3 & 1.18$\times$10$^9$ & 26.1\%\\
  BC4 (baseline) & 9.34$\times$10$^8$ & N/A\\
  BC5 & 1.15$\times$10$^9$ & 22.6\%\\
   \hline
  \end{tabular}
\end{table}
\begin{table}[h]
  \centering
  \caption{Effect of boundary conditions on maximum bulk plasma density for the axisymmetric plasma; the lateral wall used BC4 and the electrode boundary condition was varied. }
  \label{tab:BC_neElectrode}
  \begin{tabular}{|c|c|c|}
   \hline
   Boundary & Bulk plasma & \% difference \\
   condition & density (cm$^{-3}$) & (vs BC4) \\
   \hline
  BC1 &  1.45$\times$10$^9$ & -8.7\%\\
  BC2 &  3.89$\times$10$^9$ & 145.3\%\\
  BC3 & 1.90$\times$10$^9$ & 19.6\%\\
  BC4 (baseline) & 1.58$\times$10$^9$ & N/A\\
  BC5 & 1.96$\times$10$^9$ & 23.5\%\\
   \hline
  \end{tabular}
\end{table}
\begin{table}[h]
  \centering
  \caption{Effect of boundary conditions on maximum bulk plasma density for the axisymmetric plasma; the electrodes used BC4 and the wall boundary condition was varied.}
  \label{tab:BC_neWall}
  \begin{tabular}{|c|c|c|}
   \hline
   Boundary & Bulk plasma & \% difference \\
   condition & density (cm$^{-3}$) & (vs BC4) \\
   \hline
  BC1 &  1.60$\times$10$^9$ & 0.8\%\\
  BC2 &  9.27$\times$10$^8$ & -41.5\%\\
  BC3 & 1.56$\times$10$^9$ & -1.5\%\\
  BC4 (baseline) & 1.58$\times$10$^9$ & N/A\\
  BC5 & 1.57$\times$10$^9$ & -0.9\%\\
   \hline
  \end{tabular}
\end{table}
\clearpage

\end{document}